\documentclass[prd,preprint,superscriptaddress,preprintnumbers,eqsecnum,showpacs,nofootinbib,nobibnotes,noeprint]{revtex4-1}
\usepackage[linkcolor=blue,citecolor=blue,urlcolor=blue,colorlinks=true,breaklinks]{hyperref} 

\usepackage{amsfonts,bm}
\usepackage{amsfonts,amssymb,amsmath}
\usepackage{xfrac}
\usepackage{array}
\usepackage{makecell}
\usepackage{multirow}
\usepackage{graphicx}
\usepackage{paralist}
\usepackage{enumitem}
\usepackage{float}
\usepackage[english]{babel}
\usepackage{slashed}
\usepackage[usenames]{xcolor}
\usepackage{mathtools}
\usepackage[makeroom]{cancel}
\usepackage{color}

\usepackage[mathscr,scaled=1.15]{urwchancal}

\def\srm#1{{\rm{\scriptscriptstyle #1}}}

\newcommand{\be}{\begin{equation}}
\newcommand{\bea}{\begin{eqnarray}}
\newcommand{\ee}{\end{equation}}
\newcommand{\eea}{\end{eqnarray}}

\def\1eq#1{Eq.~(\ref{#1})}

\def\2eqs#1#2{Eqs.~(\ref{#1}) and~(\ref{#2})}
\def\3eqs#1#2#3{Eqs.~(\ref{#1}),~(\ref{#2}) and~(\ref{#3})}

\def\fig#1{Fig.~\ref{#1}}

\def\ie{{\it i.e.}, }
\def\eg{{\it e.g.}, }

\newcommand{\Ls}{ \mathit{L}_{{sg}}}   
\def\g{\Gamma}

\def\s#1{{\scriptscriptstyle #1}}

\newcommand{\fatg}{{\rm{I}}\!\Gamma}

\newcommand{\fatgb}{\overline{\fatg}\vphantom{\fatg}}

\newcommand{\Cfat}{{\mathbb C}}

\newcommand{\Qfat}{{\mathbb Q}}

\newcommand{\Rc}[1]{ V_{#1} }

\def\MOMt{$\widetilde{\text{MOM}}$}

\newcommand{\uf}{U}

\begin{document}


\title{
Schwinger displacement of the 
quark-gluon vertex}

\author{A.~C. Aguilar}
\affiliation{\mbox{University of Campinas - UNICAMP, Institute of Physics Gleb Wataghin,} \\
13083-859 Campinas, S\~{a}o Paulo, Brazil}

\author{M.~N. Ferreira}
\affiliation{\mbox{Department of Theoretical Physics and IFIC, 
University of Valencia and CSIC},
E-46100, Valencia, Spain}

\author{D. Iba\~nez}
\affiliation{\mbox{University Centre EDEM}, Muelle de la Aduana, La Marina de Valencia, 46024, Valencia, Spain.}

\author{J. Papavassiliou}
\affiliation{\mbox{Department of Theoretical Physics and IFIC, 
University of Valencia and CSIC},
E-46100, Valencia, Spain}

\begin{abstract}
The action of the 
Schwinger mechanism in pure Yang-Mills theories 
endows gluons with an effective mass, 
and, at the same time, 
induces a measurable 
displacement to the Ward identity
satisfied by 
the three-gluon vertex. 
In the present work we turn to 
Quantum Chromodynamics
with two light quark flavors, 
and explore the appearance  
of this characteristic displacement   
at the level of the quark-gluon vertex.
When the Schwinger mechanism is activated, 
this vertex acquires massless 
poles, 
whose momentum-dependent residues  
are determined 
by a set of coupled integral equations.
The main effect of these residues is to 
displace the Ward identity 
obeyed by the pole-free part 
of the vertex, 
causing modifications to  
its form factors, and especially 
the one associated with the tree-level tensor.
The comparison between the 
available lattice data for this  
form factor
and the Ward identity prediction 
reveals  a marked deviation,  
which is completely compatible 
with the theoretical expectation 
for the attendant residue.
This analysis corroborates further 
the self-consistency of this mass-generating scenario in the general context of real-world 
strong interactions. 
\end{abstract}

\maketitle

\section{Introduction}\label{sec:intro}

In recent years, 
the generation of 
a mass scale in the gauge sector 
of Quantum Chromodynamics (QCD)~\cite{Marciano:1977su} through the action 
of the Schwinger mechanism (SM)~\cite{Schwinger:1962tn,Schwinger:1962tp}
has emerged as a 
particularly appealing 
and robust notion~\cite{Smit:1974je,Eichten:1974et,Cornwall:1981zr,Aguilar:2011xe,Binosi:2012sj,Aguilar:2016vin,Eichmann:2021zuv,Aguilar:2021uwa,Aguilar:2022thg,Papavassiliou:2022wrb,Ferreira:2023fva}; for alternative approaches see~\cite{Fischer:2008uz,Wilson:1994fk,Kondo:2001nq,Epple:2007ut,Campagnari:2010wc,Tissier:2010ts,Alkofer:2011pe,Serreau:2012cg,Siringo:2015wtx,Cyrol:2016tym,Glazek:2017rwe,Huber:2020keu,Horak:2022aqx}. 
In its most general formulation, the SM
is based on the key  
observation that, 
even if a mass is 
forbidden at the level of the fundamental Lagrangian,
a gauge boson may become massive if its vacuum polarization 
develops  
a pole at zero momentum transfer (``{\it massless pole}'').
Such massless poles may occur for a variety 
of reasons, depending on the 
dimensionality of space-time 
and the dynamical details of each 
theory~\cite{Zumino:1965rka,Jackiw:1973tr,Jackiw:1973ha,Cornwall:1973ts}.

The non-Abelian version of the SM operating in Yang-Mills theories 
proceeds via the 
inclusion of 
massless scalar excitations  
in the fundamental 
vertices of the theory~\mbox{\cite{Smit:1974je,Eichten:1974et,Poggio:1974qs,Cornwall:1979hz,
Cornwall:1981zr,Aguilar:2008xm,Aguilar:2011xe}}, thus 
altering profoundly their analytic properties. 
These excitations 
are longitudinally coupled, and 
arise dynamically, as  
color-carrying bound states
of two gluons or a ghost-antighost pair~\cite{Aguilar:2011xe,Papavassiliou:2022wrb,Binosi:2017rwj,Aguilar:2022thg,Ferreira:2023fva,Aguilar:2021uwa,Aguilar:2017dco,Eichmann:2021zuv}.
Thanks to the functional equations 
that couple 
propagators and vertices~\cite{Roberts:1994dr,Cloet:2013jya,Alkofer:2000wg,Fischer:2006ub,Roberts:2007ji,Binosi:2009qm,Aguilar:2015bud,Huber:2018ned,Aguilar:2022thg,Papavassiliou:2022wrb,Ferreira:2023fva,Fischer:2008uz,Eichmann:2021zuv,Aguilar:2021uwa}, these  
structures 
make their way to  
the gluon self-energy, 
providing the kinematic basis 
for the appearance of a mass~\mbox{\cite{Aguilar:2011xe,Aguilar:2017dco,Binosi:2017rwj}}. 

Especially 
important in this 
context is the 
residue 
function
associated with the 
pole of the 
three-gluon vertex, denoted by~$\Cfat(r^2)$~\cite{Aguilar:2021uwa,Aguilar:2022thg,Papavassiliou:2022wrb,Ferreira:2023fva}. 
This function acts as the 
Bethe-Salpeter (BS) amplitude that 
controls the formation of the 
aforementioned bound states out of a pair of gluons~\mbox{\cite{Aguilar:2011xe,Aguilar:2017dco,Binosi:2017rwj}}.
In addition, it 
is intimately connected 
with a characteristic 
displacement of the Ward identity (WI) satisfied by the pole-free part of the three-gluon 
vertex, namely the component measured on the 
lattice~\mbox{\cite{Athenodorou:2016oyh,Duarte:2016ieu,Boucaud:2017obn,Pinto-Gomez:2022brg,Aguilar:2019uob,Aguilar:2021lke,Aguilar:2021okw}}.
In particular,
the form factor 
of the three-gluon vertex associated with 
the so-called ``soft-gluon limit'' is shown to deviate 
from the WI prediction – the correct result when the SM is inactive –
precisely by an amount $\Cfat(r^2)$.
Most importantly, 
as was established  in~\cite{Aguilar:2022thg,Aguilar:2021uwa},
this displacement is 
clearly observed in the 
lattice data: the 
$\Cfat(r^2)$ reconstructed 
is unequivocally  nonvanishing,
and in excellent agreement with the 
BS prediction~\cite{Aguilar:2021uwa}.

In the present work we consider for the first time 
the appearance 
of an analogous effect at the level of 
the fully-dressed quark-gluon vertex, thus implementing the pending generalization of the 
SM from quarkless QCD 
(pure Yang-Mills theory) 
to the case of QCD with 
two light 
quark flavors ($N_f=2$).
We emphasize that the quark-gluon vertex is 
a crucial 
component of the 
QCD dynamics, and 
has been 
studied extensively from the 
perturbative point of view~\cite{Ball:1980ay,Kizilersu:1995iz,Davydychev:2000rt,Gracey:2014mpa,Gracey:2011vw,Bermudez:2017bpx}, 
by means of continuous non-perturbative methods~\cite{Bender:1996bb,Bhagwat:2004kj,LlanesEstrada:2004jz,Holl:2004qn,Matevosyan:2006bk,Fischer:2006ub,Aguilar:2010cn,Qin:2013mta,Binosi:2016wcx,
Hopfer:2013np,Aguilar:2013ac,Rojas:2013tza,Williams:2014iea,Williams:2015cvx,Sanchis-Alepuz:2015qra,Pelaez:2015tba,Alkofer:2008tt,Mitter:2014wpa,Cyrol:2017ewj,Aguilar:2014lha,Gao:2021wun,Binosi:2016wcx,Aguilar:2016lbe,Oliveira:2018ukh,Albino:2018ncl,Tang:2019zbk,Albino:2021rvj}, 
and through numerous lattice simulations~\cite{Skullerud:2002sk,Skullerud:2002ge,Skullerud:2003qu,Skullerud:2004gp,Lin:2005zd,Kizilersu:2021jen,Kizilersu:2006et,Sternbeck:2017ntv,Skullerud:2021pel,Oliveira:2016muq,Oliveira:2018fkj}. 

The central result of the present work may be summarized by stating 
that the action of the SM 
endows the quark-gluon vertex with 
a nontrivial pole content.
The general pole structure 
admitted by Lorentz invariance and charge conjugation symmetry 
is comprised by three 
form factors, 
which, in the soft-gluon limit, are the 
quark-gluon vertex 
counterparts of the  $\Cfat(r^2)$.
In this limit, 
the essential  dynamics 
are described by a set 
of coupled BS equations, which 
account for the fact 
that the 
inclusion of active quarks permits the 
additional formation of composite 
scalars 
out of quark-antiquark 
pairs. 

The numerical treatment of these equations demonstrates  the nonvanishing 
nature of the residue functions 
associated with the poles of the quark-gluon vertex. 
Moreover, the 
modifications produced by their presence 
to the unquenched version of 
$\Cfat(r^2)$ are rather minimal, 
and the mass-generating aspects 
of the SM remain practically unaltered.
These findings clearly validate 
the expectation~\cite{Aguilar:2022thg} that the gluon mass generation in the 
presence of dynamical quarks is 
qualitatively similar to the 
pure Yang-Mills case. In fact, our preliminary analysis 
is  
compatible with the lattice results of 
\cite{Ayala:2012pb,Binosi:2016xxu}, which indicate that the unquenched gluon mass is larger than the quenched one.

Interestingly enough, however, the most important effect of these novel residue functions is the distinctive displacement that they produce 
to the WI satisfied by the 
pole-free part of the quark-gluon vertex.
Focusing on the form factor 
associated with the classical tensorial structure, $\gamma_{\alpha}$, this displacement introduces a sharp 
difference between the result 
computed on the lattice, $\lambda_1(p^2)$,
and the prediction of the WI in the 
absence of the SM, $\lambda_1^{\star}(p^2)$.

The above predictions may be tested 
by appealing to the available  
lattice data for this
form factor~\cite{Kizilersu:2021jen}. In particular, we focus on the two asymmetric lattices, 
denominated ``L08'' and ``L07'' 
in~\cite{Kizilersu:2021jen}, 
which have superior statistics and   
small current quark masses. 
The detailed comparison between 
$\lambda_1(p^2)$ and $\lambda_1^{\star}(p^2)$
produces a 
clear signal 
of 
$8\sigma$ and 
$6\sigma$, respectively,
as shown in \fig{fig:lambda1}.
In fact, the resulting curves for the 
displacement function 
are in  
very good 
agreement with the BS prediction. 
We consider these novel results as  
an additional 
indication of the operation of the SM 
in QCD. 

The article is organized as follows. 
In Sec.~\ref{sec:gen} we review the most salient features of the SM in the context of a Yang-Mills
theory, focusing on the 
displacement function associated with the 
three-gluon vertex, which serves as the prototype 
for the ensuing construction. 
In Sec.~\ref{sec:quark_disp}
we demonstrate how the action of the SM 
modifies the WI prediction 
for the 
form factors of 
the quark-gluon vertex 
in the soft-gluon limit, and introduce the 
corresponding displacement functions.
In Sec.~\ref{sec:bse} we solve the 
system of coupled BS equations that 
determines the set of displacement functions  
associated with the three-gluon and quark-gluon 
vertices.  
Then, in Sec.~\ref{sec:lat_disp}
we present the central result of the present 
work. In particular, we determine the 
displacement function associated with 
the tree-level component of the quark-gluon vertex  
from the 
difference between the 
lattice data and the WI prediction.
In Sec.~\ref{sec:conc} we present our discussion and conclusions. 
Finally, in three Appendices we provide plots and numerical fits for the lattice inputs used in our computations, explain the implementation of the necessary transitions between the three main 
renormalization schemes employed in the literature,
and give some technical details on the Schwinger-Dyson equation (SDE) determination 
of a special form factor.
    
\section{Schwinger mechanism and Ward identity displacement}\label{sec:gen} 

In this section we review the main concepts of the 
WI  displacement 
of the 
three-gluon vertex, which is induced 
by the action of the SM.
These  
considerations set the stage for accomplishing the main objective of this
 work, namely 
 study of the same effect 
 in the case of the quark-gluon vertex. 

\begin{figure}[ht]
\centering
\includegraphics[width=0.75\linewidth]{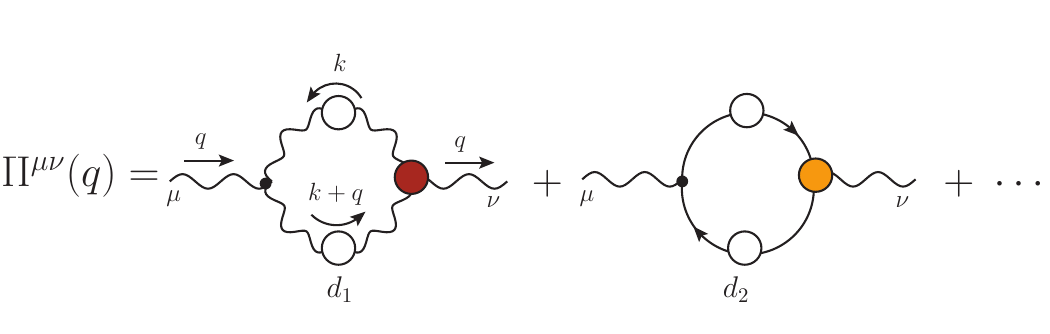}
\caption{ Diagrammatic representation of the fully-dressed unquenched gluon self-energy, $\Pi_{\mu\nu}(q)$, entering in the definition of the SDE for the gluon propagator, given by Eq.~\eqref{glSDE}.
White (colored) circles indicate fully-dressed propagators (vertices).}
\label{fig:pi_unqueched}
\end{figure}

 In the {\it Landau gauge} that we use throughout this work, 
the gluon propagator, $\Delta^{ab}_{\mu\nu}(q)$, has the form\footnote{We work in Minkowski space, passing key results 
to Euclidean space for the purposes of numerical 
evaluation or comparison with the lattice.} 
$\Delta^{ab}_{\mu\nu}(q)=-i\delta^{ab}\Delta_{\mu\nu}(q)$,
 with 
\be
\Delta_{\mu\nu}(q) = \Delta(q^2) {P}_{\mu\nu}(q)\,, \qquad {P}_{\mu\nu}(q) := g_{\mu\nu} - q_\mu q_\nu/{q^2}\,.
\label{defgl}
\ee
%
The momentum evolution of  
$\Delta(q^2)$ in the continuum is determined by
the corresponding SDE, given by 
\be
\Delta^{-1}(q^2)P_{\mu\nu}(q) = q^2P_{\mu\nu}(q)  + i \,{\Pi}_{\mu\nu}(q) \,,
\label{glSDE}
\ee
where $\Pi_{\mu\nu}(q) = q^2 {\bf \Pi}(q^2) P_{\mu\nu}(q)$ 
is the transverse gluon self-energy,
shown diagrammatically in \fig{fig:pi_unqueched}.

A pivotal component 
of this SDE is 
the fully-dressed three-gluon vertex, 
$\fatg^{abc}_{\alpha\mu\nu}(q,r,k)$, 
which may be cast in the form 
\be
\fatg^{abc}_{\alpha\mu\nu}(q,r,k)  = g f^{abc} \fatg_{\alpha\mu\nu}(q,r,k) \,,
\label{verd}
\ee
where $g$ is the gauge coupling, $f^{abc}$ are the SU(3) structure constants, and 
\mbox{$q+r+k = 0$}. 

\begin{figure}[t]
\centering
\includegraphics[width=0.9\linewidth]{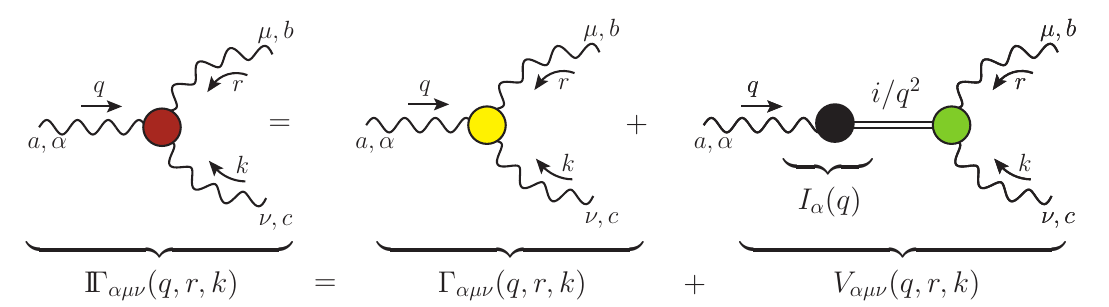}
\includegraphics[width=0.9\linewidth]{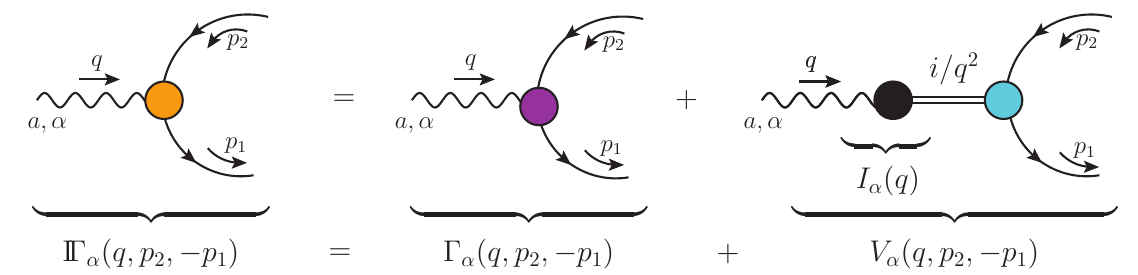}
\caption{ Diagrammatic representation 
of \1eq{fvert} (top row), 
and of \1eq{fvertq} (bottom row).
The black circle indicates the 
gluon-scalar transition function.
} 
\label{fig:split_poles}
\end{figure}

It turns out that 
the infrared saturation of the gluon propagator  
hinges crucially on the special analytic structure of $\fatg_{\alpha\mu\nu}$. 
To appreciate this point, 
recall that, according 
to the SM, if ${\bf \Pi}(q^2)$
acquires a pole with a positive residue 
(Euclidean space), 
\mbox{$\lim_{q^2 \to 0} {\bf \Pi}(q^2) = m^2/q^2$}, then  
\mbox{$\Delta^{-1}(0) = m^2$}, \ie the gluon propagator 
picks up a mass~\cite{Schwinger:1962tn,Schwinger:1962tp, Eichten:1974et,Smit:1974je,Poggio:1974qs,Cornwall:1979hz,Cornwall:1981zr, Jackiw:1973tr,Jackiw:1973ha,Cornwall:1973ts,Aguilar:2008xm,Aguilar:2021uwa,Aguilar:2022thg,Ferreira:2023fva,Papavassiliou:2022wrb}.
$\fatg_{\alpha\mu\nu}$
provides precisely this 
pole\footnote{The 
ghost-gluon and four-gluon vertices also develop poles, 
but their influence 
on the issues studied here is rather limited~\cite{Aguilar:2017dco,Aguilar:2021uwa}.},
as a result of 
specific nonperturbative dynamics: 
composite scalar excitations are formed 
(\eg out of a pair of gluons), 
which carry color, 
are longitudinally coupled, and have vanishing mass~\cite{Eichten:1974et,Smit:1974je,Poggio:1974qs,Cornwall:1979hz,Cornwall:1981zr,Aguilar:2008xm,Aguilar:2011xe,Papavassiliou:2022wrb,Aguilar:2021uwa,Aguilar:2022thg}.
As a result, $\fatg_{\alpha\mu\nu}$ may be cast in the form 
(see upper row of  \fig{fig:split_poles})
\be
\fatg_{\alpha\mu\nu}(q,r,k) = \g_{\alpha\mu\nu}(q,r,k) + 
V_{\alpha\mu\nu}(q,r,k) \,,
\label{fvert}
\ee
where $\g_{\alpha\mu\nu}(q,r,k)$ denotes the pole-free part,
while $V_{\alpha\mu\nu}(q,r,k)$ 
contains longitudinal poles 
of the type $q^{\alpha}/q^2$,
$r^{\mu}/r^2$,   
$k^{\nu}/k^2$, and products 
thereof~\cite{Aguilar:2023mdv}. 
The longitudinal
nature of $V_{\alpha\mu\nu}(q,r,k)$  is a direct consequence of the general form of the gluon-scalar
transition amplitude $I_{\alpha} (q)$
[see \fig{fig:split_poles}], 
namely 
$I_{\alpha} (q) = q_{\alpha} 
I(q^2)$, imposed by 
Lorentz symmetry~\cite{Aguilar:2011xe}; it leads 
to the important property 
${P}_{\alpha'}^{\alpha}(q){P}_{\mu'}^{\mu}(r){P}_{\nu'}^{\nu}(k) V_{\alpha\mu\nu}(q,r,k) = 0$.
This last relation 
eliminates the poles 
from lattice simulations 
of $\fatg_{\alpha\mu\nu}$
in the Landau gauge;
nonetheless, characteristic {\it finite} 
effects survive, which manifest themselves as 
displacements of certain key quantities.

Focusing on the leading pole 
in the $q$-channel, we have 
\be
V_{\alpha\mu\nu}(q,r,k) =  \frac{q_\alpha}{q^2}  g_{\mu\nu} V_{1}(q,r,k) + \cdots  \,,
\label{Vtens}
\ee
where 
the ellipsis absorbs 
all remaining  
pole contributions,
which 
get annihilated 
upon the contraction by the 
projectors 
$P^{\mu'\mu}(r)$
or 
$P^{\nu'\nu}(k)$.

The kinematic limit  
relevant in this study 
is that of $q\to 0$. 
A detailed analysis~\cite{Aguilar:2016vin}  
shows that $V_1(0,r,-r) = 0$, and 
the Taylor expansion of $\Rc1(q,r,k)$ around $q= 0$ yields
\be 
\label{eq:taylor_C}
	\lim_{q \to 0} \Rc1(q,r,k) = 2 (q\cdot r) \, \Cfat(r^2)
+ {\cal O}(q^2)\,,
\ee
where 
$\Cfat(r^2) := 
[\partial \Rc1(q,r,k)/\partial k^2]_{q = 0}$.

The function $\Cfat(r^2)$
is of central importance in this 
approach, playing three key
roles:  

({\it i})
$\Cfat(r^2)$ is  
the {\it BS amplitude} describing the bound-state formation of a massless  
{\it colored} scalar out of a pair 
of gluons. In particular,  $\Cfat(r^2)$
is obtained as the 
nontrivial solution 
of an appropriate  
linear and homogeneous 
BS equation,
deduced from the SDE of $\fatg_{\alpha\mu\nu}$
in the 
limit $q\to 0$.

({\it ii}) 
The pole $1/q^2$ 
appearing in \1eq{Vtens}
is transmitted to ${\bf \Pi}(q^2)$ 
via the gluon SDE of \1eq{glSDE}, 
triggering the SM. 
The gluon mass is then determined by integrals involving 
$\Cfat(r^2)$, of the general form\footnote{ 
In terms of the gluon-scalar transition amplitude 
given by the exact relation $m^2 = g^2 I^2(0)$~\cite{Smit:1974je,Eichten:1974et,Ibanez:2012zk}, where $I(0)$ is expressed in terms of integrals such as the one on the r.h.s. of \1eq{Dgl}.}
\be
m^2 \sim \int\!\! d^4 k \, k^2 \Delta^2(k^2)  \,\Cfat(k^2)\,.
\label{Dgl}
\ee

({\it iii})
If we define the projector
${\cal T}_{\mu'\nu'}^{\mu\nu}(r,k) :=
P_{\mu'}^{\mu}(r)P_{\nu'}^{\nu}(k)$,
the  Slavnov-Taylor identity (STI)~\cite{Taylor:1971ff,Slavnov:1972fg} satisfied by 
$\fatg_{\alpha\mu\nu}(q,r,k)$ may be cast in the form
\be
q^\alpha \fatg_{\alpha \mu \nu}(q,r,k)
\,{\cal T}_{\mu'\nu'}^{\mu\nu}(r,k)
\!=\! F(q^2)\,[ 
 \Delta^{-1}(k^2) H_{\nu\mu}(k,q,r) \!-\! \Delta^{-1}(r^2) H_{\mu\nu}(r,q,k)]
\,{\cal T}_{\mu'\nu'}^{\mu\nu}(r,k)\,,
\label{STI} 
\ee
where $F(q^2)$
is the ghost dressing function, and 
$H_{\nu\mu}(k,q,r)$ the 
ghost-gluon kernel. 
Note that, 
by virtue of 
\2eqs{fvert}{Vtens}, 
the l.h.s. of \1eq{STI} becomes 
\be
q^\alpha \fatg_{\alpha \mu \nu}(q,r,k)
\,{\cal T}_{\mu'\nu'}^{\mu\nu}(r,k)=
[q^\alpha \g_{\alpha\mu\nu}(q,r,k) + 
g_{\mu\nu} V_{1}(q,r,k)]
{\cal T}_{\mu'\nu'}^{\mu\nu}(r,k)\,.
\ee
When the limit 
$q\to 0$ 
of both sides 
of \1eq{STI} 
is taken, 
one obtains the 
corresponding WI, which is  
tantamount to the 
so-called 
``soft-gluon limit'' of the three-gluon vertex.
In doing so, 
\1eq{eq:taylor_C} 
is triggered 
and $\Cfat(r^2)$ 
makes its appearance. 
Moreover, 
as has been shown in detail in~\cite{Aguilar:2021okw}, 
\be
{\cal T}_{\mu'\nu'}^{\mu\nu}(r,-r)\Gamma_{\alpha\mu\nu}(0,r,-r) =   2 r_{\alpha} 
P_{\mu'\nu'}(r) \Ls(r^2) \,,
\label{Lsg}
\ee
where $\Ls(r^2)$ is precisely the form factor 
measured on the lattice through the ratio
\be 
\Ls(r) = \left. \frac{\g_0^{\alpha\mu\nu}(q,r,k)P_{\alpha\alpha'}(q)P_{\mu\mu'}(r)P_{\nu\nu'}(k)\fatg^{\alpha'\mu'\nu'}(q,r,k)}{\g_0^{\alpha\mu\nu}(q,r,k)P_{\alpha\alpha'}(q)P_{\mu\mu'}(r)P_{\nu\nu'}(k)\g_0^{\alpha'\mu'\nu'}(q,r,k)} \right\vert_{q\to 0} \,, \label{Lsg_def}
\ee
where $\g_0^{\alpha\mu\nu}$ is the tree-level expression of the three-gluon vertex, given by 
\be
\Gamma_{\!0}^{\alpha\mu\nu}(q,r,k)  = (q-r)^{\nu}g^{\alpha\mu} + (r-k)^{\alpha}g^{\mu\nu} + (k-q)^{\mu}g^{\alpha\nu}\,. 
\label{3g_tree}
\ee

The final upshot of these considerations is that the function $\Cfat(r^2)$
leads to the displacement
of the soft-gluon limit
according to 
(Euclidean space) 
\be
\underbrace{\Ls(r^2)}_{
{\rm SM \, on}} \, = \, \underbrace{\Ls^{\!\star}(r^2)}_{\rm SM \, off} \,\, +  \underbrace{\Cfat(r^2)}_{\substack{{\rm displacement} \\ {\rm function} } }\,,
\label{3gldis}
\ee
where 
\be 
\Ls^{\!\star}(r^2) = 
[{\rm r.h.s. \,\, of \, \,STI \,\,in \,\, \1eq{STI}}]_{q\to 0}\,, 
\label{Lstar}
\ee 
is
the theoretical predictions for the same form factor when the 
SM is turned off.

Most importantly, 
as has been established recently in~\cite{Aguilar:2022thg},  
the difference between  
 $\Ls(r^2)$
and $\Ls^{\!\star}(r^2)$
corresponds to 
a clearly nonvanishing 
$\Cfat(r^2)$, whose shape 
is absolutely compatible with that 
obtained in ({\it i}).

\begin{figure}[t]
\centering
\begin{minipage}[c]{.45\linewidth}
\includegraphics[width=1\linewidth]{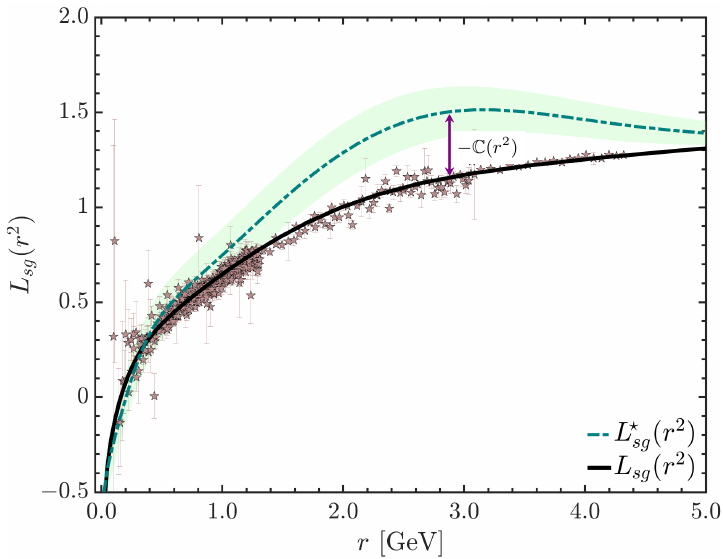}
\end{minipage}
\hfil
\begin{minipage}[c]{.45\linewidth}
\includegraphics[width=1\linewidth]{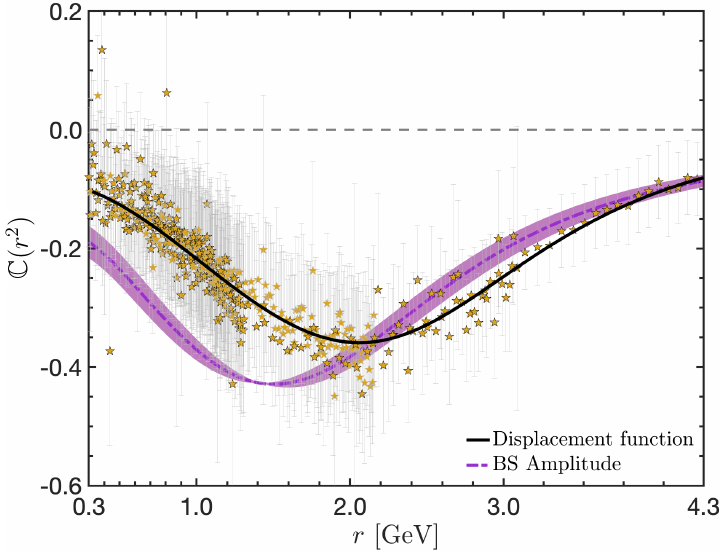}
\end{minipage}
\caption{ Left panel: Quenched lattice data for $\Ls(r^2)$ from \cite{Aguilar:2021lke} (points + black continuous curve) compared to the $\Ls^{\!\star}(r^2)$ (green dashed) determined in \cite{Aguilar:2022thg}. The green band represents the error in $\Ls^{\!\star}(r^2)$, estimated as explained in detail in~\cite{Aguilar:2022thg}. Right panel: The displacement function, $\Cfat(r^2)$, obtained from the results on the left panel through \1eq{3gldis} (points + black continuous 
curve), compared to the BS amplitude of~\cite{Aguilar:2021uwa} (purple line and corresponding error band).  } 
\label{fig:Lsg_disp}
\end{figure}

As we will see in the
next section, 
the SM produces 
a displacement 
analogous to that of \1eq{3gldis}
in the form factors of the quark-gluon vertex.

\begin{table}[H]
\begin{tabular}{|c|c|c|c|}
\hline
\multicolumn{4}{|c|}{$\Cfat(r^2)$} \\
\hline
\makecell{Defining \\ equation} & $\displaystyle{\lim_{q\to 0}} \frac{V_1(q,r,k)}{q^2} = \frac{2 (q\cdot r)}{q^2}\Cfat(r^2)$ & $\Cfat(r^2) = \int \!\small{d}^{\s{4}}\!\small{\ell}\, {\cal K}(r,\ell) \Cfat(\ell^2)$ & $\Ls(r^2) = \Ls^{\!\star}(r^2) + \Cfat(r^2)$ \\
\hline
Terminology & \makecell{Momentum-dependent residue \\ or \\ residue function } & BS amplitude & Displacement function\\
\hline
\end{tabular}
\caption{ Equivalent definitions of the function $\Cfat(r^2)$ and the corresponding terminology. \label{tab:Cfat}}
\end{table}

We close this introductory section by pointing 
out that, due to its multiple roles, 
the function $\Cfat(r^2)$ receives  
a variety of names, which, depending on the situation, emphasize different physical aspects; 
the terminology employed 
and its origin is summarized in  Table \ref{tab:Cfat}.
Completely analogous terminology is employed for the 
corresponding quantities associated with the quark-gluon vertex,
introduced in the next section. 

\section{Quark-gluon vertex and displaced soft-gluon limit}\label{sec:quark_disp} 

The previous considerations have been carried out in the realm of pure Yang-Mills theories, 
or quarkless QCD.  
We now turn to the main part of this study, 
and consider a real-world version of 
QCD, with two light quark flavors. 

In this case, the main modification with respect to the 
Yang-Mills case is the inclusion 
of 
the fully-dressed quark-gluon vertex, 
$\fatg^{a}_{\alpha}(q,p_2,-p_1)$,
which may be cast in the form 
\be
\fatg_\alpha^a(q,p_2,-p_1) = ig \,\frac{\lambda^a}{2}\,\fatg_\alpha(q,p_2,-p_1) \,, 
\ee
where 
the $\lambda^a$ denote the 
standard Gell-Mann matrices, and the four-momenta 
satisfy the relation $q+p_2-p_1=0$. 

At tree-level, $\fatg_\alpha(q,p_2,-p_1)$ acquires the standard expression
\be
\g_{\!0}^\alpha(q,p_2,-p_1) = \gamma^{\alpha}\,.
\label{qg-tree}
\ee

The activation of the SM
generates longitudinal poles in the only channel 
that carries a Lorentz index, namely the 
$q$-channel. 
Then, as shown in the bottom row of \fig{fig:split_poles}, the quark-gluon vertex is 
composed by two distinct 
pieces, 
\be
\fatg_\alpha(q,p_2,-p_1) = \g_\alpha(q,p_2,-p_1) + V_\alpha(q,p_2,-p_1)\,,
\label{fvertq}
\ee
where $\g_\alpha(q,p_2,-p_1)$ is the pole-free part, and 
\be 
V_\alpha(q,p_2,-p_1) = \frac{q_\alpha}{q^2}Q(q,p_2,-p_1) \,,
\label{eq:Vquark}
\ee
is the pole term.
 
The amplitude $Q(q,p_2,-p_1)$ 
may be decomposed 
in 
a standard Dirac basis according to  
\begin{equation}
\label{CQdecomposition}
Q(q,p_2,-p_1) = Q_1\,\mathbb{I} + Q_2\,\slashed{p}_2 + Q_3\,\slashed{p}_1 + Q_4\,\tilde{\sigma}_{\mu\nu}\, p_2^\mu \, p_1^\nu \,,
\end{equation} 
where $\mathbb{I}$ is the  $4\times 4$ identity matrix in the Dirac space, \mbox{$\tilde{\sigma}^{\mu\nu} =1/2[\gamma^\mu,\gamma^\nu]$}, and we use the short-hand notation 
$Q_i \equiv Q_i(q^2,p_2^2,p_1^2)$.

\subsection{Schwinger mechanism turned off}\label{SMoff}

\begin{figure}[ht]
\includegraphics[width=0.5\linewidth,trim={0 0 0 1cm},clip]{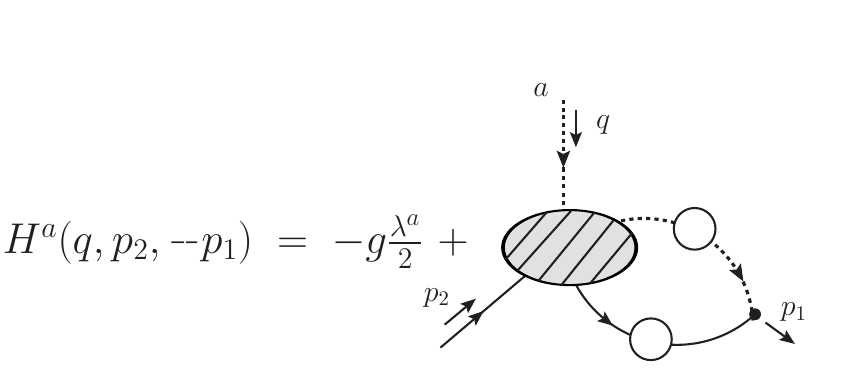} \\
\includegraphics[width=0.7\linewidth,trim={0 0.1cm 0 0},clip]{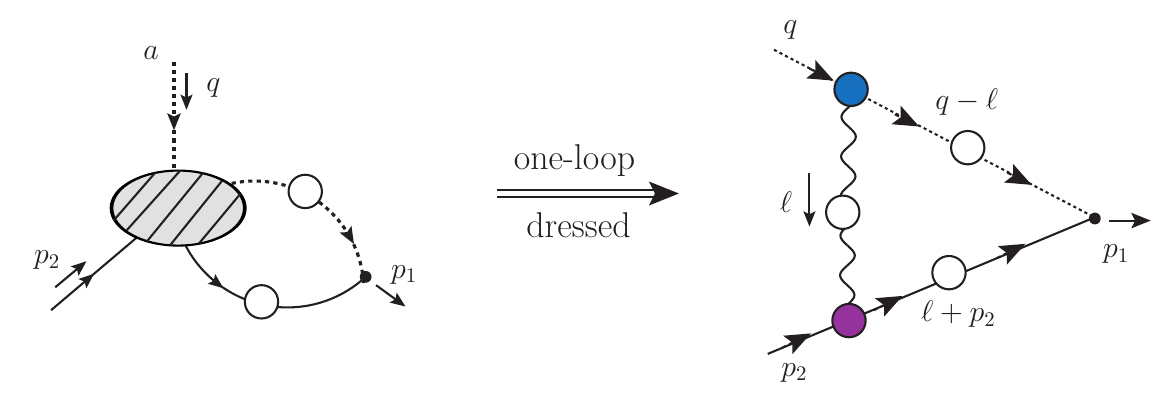}
\caption{Top panel:
Diagrammatic representation of the quark-ghost kernel, $H^a(q,p_2,-p_1)$. 
Bottom panel: 
The one-loop dressed approximation 
of $H^a(q,p_2,-p_1)$, 
employed 
for the 
determination of 
the $K_i(p^2)$ in 
Sec.~\ref{sec:lat_disp}.}
\label{H_and_K}
\end{figure}

Let us suppose that 
the SM is 
not active, such that 
$V_\alpha(q,p_2,-p_1)=0$; 
then, from \1eq{fvertq}
we have that 
$\fatg_\alpha(q,p_2,-p_1) = \g_\alpha(q,p_2,-p_1)$, \ie the full vertex is 
pole-free. Then, 
consider the STI 
triggered when 
the quark-gluon vertex 
$\g_\alpha(q,p_2,-p_1)$ 
is contracted by the 
gluon momentum $q^\alpha$, 
namely~\cite{Roberts:1994dr,Cloet:2013jya} 
\begin{equation}
\label{qfreegnp}
q^\alpha\g_\alpha(q,p_2,-p_1) = F(q^2)[S^{-1}(p_1)H(q,p_2,-p_1) - \overline{H}(-q,p_1,-p_2) \, S^{-1}(p_2)] \,,
\end{equation}
where 
$S^{ab}(p) = i \delta^{ab}S(p)$ is the quark 
propagator ~\cite{Natale:1996eu,Fleischer:1998dw,Franco:1998bm,Maris:2003vk,Bhagwat:2003vw,Bowman:2005vx,Furui:2006ks,Sauli:2006ba,Chang:2006bm,Bashir:2012fs,Oliveira:2018lln,Chen:2021ikl,Gao:2021wun}
\be
S^{-1}(p) = A(p^2)\,\slashed{p} - B(p^2)\,.
\label{SAB}
\ee
In addition, $H(q,p_2,-p_1)$ is the 
quark-ghost scattering kernel [see upper panel of Fig.~\ref{H_and_K}],
and 
$\overline{H}(-q,p_1,-p_2)$
its ``conjugate'',  whose relation to $H$ is explained in detail in~\cite{Aguilar:2016lbe} [see discussion around Eq.~(2.5)]. Note that
the color structure has been factored out, setting \mbox{$H^{a} = -g\, (\lambda^a/2) H$}. 

Next, we expand  
both sides of \1eq{qfreegnp}
around $q=0$, with $p_1 = p_2 = p$. It is clear that 
the zeroth order term vanishes on both sides; then, the matching of the 
coefficients multiplying the 
terms linear in 
$q$ yields the corresponding WI.

We start by noting that the term linear in $q$ on 
the l.h.s of \1eq{qfreegnp} 
is simply 
$q^\alpha\g_\alpha(0,p,-p)$, where 
$\g_\alpha(0,p,-p)$ is the soft-gluon limit of the 
quark-gluon vertex. The most general Lorentz decomposition of the 
vertex $\g_\alpha(0,p,-p)$
is given by  
\begin{equation}
\label{softgQG}
\g_\alpha(0,p,-p) = \Gamma_1(p^2)\,\gamma_\alpha + 4\Gamma_2(p^2)\,\slashed{p}\,p_\alpha + 2\Gamma_3(p^2)\,p_\alpha + 2\Gamma_4(p^2)\,\tilde{\sigma}_{\alpha\nu}p^\nu \,, 
\end{equation}
where $\Gamma_i(p^2)$ are the form factors in the soft-gluon limit. 
However, 
due to charge conjugation symmetry~\cite{Kizilersu:2021jen},
$\Gamma_4(p^2)$ vanishes identically, thus reducing the 
number of relevant form factors down to three.

In the Landau gauge, 
the expansion of the r.h.s. of 
 \1eq{qfreegnp} around $q=0$ is 
 facilitated by setting~\cite{Aguilar:2014lha} 
\begin{align}
H(q,p_2,-p_1) =&\, Z_H + q^\alpha K_\alpha(q,p_2,-p_1) \,,
\nonumber\\
\overline{H}(-q,p_1,-p_2) =&\, Z_H - q^\alpha \overline{K}_\alpha(-q,p_1,-p_2) \,, 
\label{LandauH}
\end{align}
where $Z_H$ is the quark-ghost kernel renormalization constant, which is finite in Landau gauge~\cite{Taylor:1971ff}. The forms of the $K_\alpha$ and $\overline{K}_\alpha$ have been given in~\cite{Aguilar:2014lha}.
Thus, the first two terms in the 
Taylor expansion of these kernels are given by  
\begin{align}
H(0,p,-p) =&\, Z_H + q^\alpha K_\alpha(0,p,-p) \,,
\nonumber\\
\overline{H}(0,p,-p) =&\, Z_H - q^\alpha\overline{K}_\alpha(0,p,-p)\,, 
\label{sgl}
\end{align}
where, employing the same basis as in 
\1eq{softgQG}, we find 
\begin{align}
K_\mu(0,p,-p) =&\, K_1(p^2)\,\gamma_\mu + 4K_2(p^2)\,\slashed{p}\,p_\mu + 2K_3(p^2)\, p_\mu + 2K_4(p^2)\,\tilde{\sigma}_{\mu\nu}p^\nu \,,
\nonumber\\
\overline{K}_\mu(0,p,-p) =&\, K_1(p^2)\,\gamma_\mu + 4K_2(p^2)\,\slashed{p}\,p_\mu + 2K_3(p^2)\, p_\mu - 2K_4(p^2)\,\tilde{\sigma}_{\mu\nu}p^\nu \,, 
\label{sgKdec}
\end{align}
with $ K_i(p^2)$ being the form factors. 
In addition, we use the following expansion 
for the inverse quark propagator,
\be
\label{invSTaylor}
S^{-1}(p_1) \!= \!S^{-1}(p) \, + \,q^\alpha\bigg[A(p^2)\gamma_\alpha + 
2A'(p^2)\,\slashed{p}\,p_\alpha - 2 B'(p^2) p_\alpha\bigg] \,,
\ee
where  
the short-hand notation $f'(p^2) := {df(p^2)}/{d p^2}$ is used. 

Then, the matching of the terms 
on the two sides of 
\1eq{qfreegnp} expresses 
the form factors $\Gamma_i(p^2)$
in terms of the functions $\{A,B,F,K_i\}$. 
Converting the results to Euclidean space 
by means of standard transformation rules, 
and setting 
\mbox{$\lambda_i^{\star}(p^2) := \Gamma_i^E(p^2_{\s E})$},
we obtain 
\begin{align}
\lambda_1^{\star}(p^2) =&\, F(0)\left\lbrace \left[Z_H + 4p^2 K_4(p^2)\right] A(p^2)  - 2K_1(p^2) B(p^2) \right\rbrace \,,
\nonumber\\
\lambda_2^{\star} (p^2) =&\, F(0)\left\lbrace - \frac{1}{2}A'(p^2)Z_H + \left[ K_3(p^2) + K_4(p^2)\right] A(p^2)  + 2K_2(p^2)  B(p^2) \right\rbrace \,,
\nonumber \\
\lambda_3^{\star} (p^2) =&\, F(0)\left\lbrace Z_H B'(p^2) + \left[ K_1(p^2) + 4p^2K_2(p^2) \right]A(p^2) - 2K_3(p^2)B(p^2)\right\rbrace \,.
\label{lamdai}
\end{align}
Note that \1eq{lamdai} coincides with the Euclidean form of Eq.~(4.3) in~\cite{Aguilar:2014lha} up to redefinitions\footnote{The $K_i$ of \cite{Aguilar:2014lha} can be obtained from the present ones by the redefinitions $K_1 \to K_1$, $K_2\to K_2/4$, $K_3\to K_3/2$ and $K_4\to K_2/2$.}. Also, note that \cite{Aguilar:2014lha} employed explicitly the Taylor renormalization scheme, where \mbox{$Z_H = 1$}~\cite{Taylor:1971ff}.

\subsection{Schwinger mechanism turned on}\label{SMon}

Let us now 
restore the 
full content of \1eq{fvertq}, \ie
allow for the presence of massless poles in 
the fundamental vertices of the theory.
With the Schwinger mechanism activated, the 
form of the STI remains intact~\cite{Smit:1974je,Aguilar:2016vin}:
the divergence of the 
full vertex, now 
\mbox{$\fatg_\alpha =\g_\alpha + V_\alpha$},
is still given by the r.h.s. of \1eq{qfreegnp}.
Since \mbox{$q^\alpha V_\alpha = Q$}, the STI satisfied by the 
pole-free part of the vertex (the one simulated on the lattice) is displaced with respect to \1eq{qfreegnp}, namely
\begin{equation}
\label{qfreeg}
q^\alpha\g_\alpha(q,p_2,-p_1) = F(q^2)[S^{-1}(p_1)H(q,p_2,-p_1) - \overline{H}(-q,p_1,-p_2) \, S^{-1}(p_2)] -
\underbrace{Q(q,p_2,-p_1)}_{\rm displacement}.
\end{equation}
Consequently, the form factors of the vertex $\g_\alpha$ 
in the presence of the Schwinger mechanism, to be denoted by 
$\lambda_i$, 
will be also displaced with respect to the $\lambda_i^{\star}$
given in \1eq{lamdai}.
To determine this displacement in detail, 
the construction leading to \1eq{lamdai} 
must be supplemented by the linear contributions
arising from 
the Taylor expansion of $Q(q,p_2,-p_1)$ around $q=0$.

To that end, note that, 
since at $q=0$ all other terms in 
\1eq{qfreeg} vanish, the condition $Q(0,p,-p) = 0$  
must be fulfilled. Substituting $q=0$ into  
\1eq{CQdecomposition}, and setting 
\mbox{$Q_i (0,p^2,p^2) :=Q_i(p^2)$}, we obtain the constraints 
\be
\label{C1qsoft}
Q_1(p^2) = Q_2(p^2) + Q_3(p^2) =0 \,. 
\ee
This argument leaves $Q_4(p^2)$ undetermined; however, charge conjugation symmetry
forces $Q_4(p^2)$ to vanish, $Q_4(p^2)=0$.

Thus, the Taylor expansion of $Q(q,p_2,-p_1)$ is given by 
\begin{equation}
\label{TaylorCq}
\lim_{q\to 0} Q(q,p_2,-p_1) = q^\alpha
\left[\frac{\partial Q(q,p_2,-p_1)}{\partial q^\alpha}\right]_{q=0} + {\cal O}(q^2) \,,
\end{equation}
and from \1eq{CQdecomposition} we find
\be
\left[\frac{\partial Q(q,p_2,-p_1)}{\partial q^\alpha}\right]_{q=0} = 2\Qfat_1(p^2)\,p_\alpha + 2\Qfat_{2+3}(p^2)\,\slashed{p}\,p_\alpha
 + Q_3(p^2)\,\gamma_\alpha \,, 
\label{derivativeTaylorCq}
\ee
where we have introduced
\begin{align}
 \Qfat_i(p^2) := &
[\partial Q(q,p_2,-p_1)/\partial p_1^2]_{q=0} \,,
\, \qquad   i=1,2,3,
\nonumber\\
\Qfat_{2+3}(p^2) :=&
\Qfat_2(p^2) + \Qfat_3(p^2) \,.
\label{derivativeCqi}
\end{align}
Then, combining \2eqs{lamdai}{TaylorCq}, we may easily determine from \1eq{qfreeg}
the displacement of the vertex form factors when the 
SM is active, namely
\begin{align}
\lambda_1(p^2) =&\,  \lambda_1^{\star}(p^2) \, - \, Q_3(p^2) \,, 
\nonumber\\
\lambda_2(p^2) =&\, \lambda_2^{\star}(p^2) \, + \,  \frac{1}{2}
\Qfat_{2+3}(p^2) \,, 
\nonumber \\
\lambda_3 (p^2) = &\, \lambda_3^{\star} (p^2) \, + \, \Qfat_1(p^2) \,.
\label{SMlamdai}
\end{align}
The above relations are the 
direct analogue of \2eqs{3gldis}{Lstar}
for the case of the quark-gluon vertex;
$Q_3(p^2)$, $\Qfat_{2+3}(p^2)$,  and $\Qfat_1(p^2)$
are the corresponding displacement functions.

We emphasize that 
the $\lambda_i(p^2)$ 
capture the full dynamical content of the theory, in the sense 
that they take into account the action of the SM. 
Therefore, if the SM is realized as described 
here, it is the $\lambda_i(p^2)$,  
and not the $\lambda_i^{\star} (p^2)$, that should coincide with the results 
obtained for these form factors from lattice QCD. Let us now suppose that the 
$\lambda_i^{\star} (p^2)$ were determined 
through \1eq{lamdai}, using lattice ingredients for all (or most) intermediate 
calculations. 
Then, if a statistically 
significant discrepancy 
is found between the $\lambda_i^{\star} (p^2)$
and the lattice $\lambda_i(p^2)$, 
it is natural to attribute it 
to the  
presence of the displacement functions. 
Such an observation becomes even more 
compelling
if the observed signal displays a 
momentum dependence compatible with the 
BS prediction for the 
corresponding 
amplitudes/displacement functions,
as happens in the case of pure Yang-Mills 
\cite{Aguilar:2022thg} 
(see \fig{fig:Lsg_disp}).
As we demonstrate in  
the rest of this article, this is indeed what happens also in the 
case of QCD with $N_f=2$.

\section{Dynamical determination of the displacement functions}\label{sec:bse}

In order to determine the numerical importance of the displacements exhibited in
\1eq{SMlamdai}, in this section we compute the displacement
functions 
from the BS equations that they obey.

\subsection{System of Bethe-Salpeter equations}\label{subsec:BSE}

\begin{figure}[h]
\centering
\includegraphics[width=0.9\linewidth]{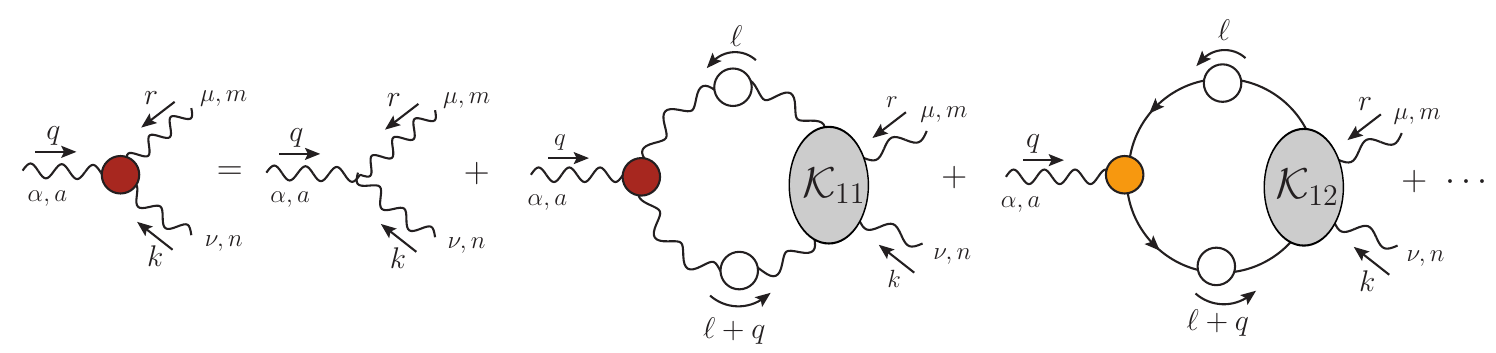} 
\includegraphics[width=0.9\linewidth]{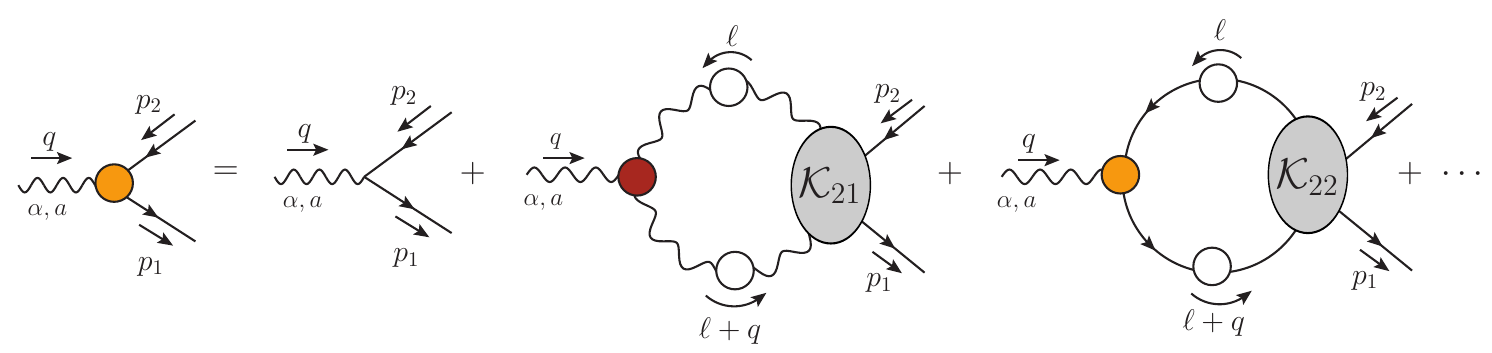}
\caption{ The
SDEs for the three-gluon (top row) and quark-gluon (bottom row) vertices. The white (colored)
circles denote fully-dressed propagators (vertices), while the gray ellipses denote four-particle
kernels.} 
\label{fig:BSE1}
\end{figure}

\begin{figure}[!t]
\includegraphics[width=0.75\linewidth]{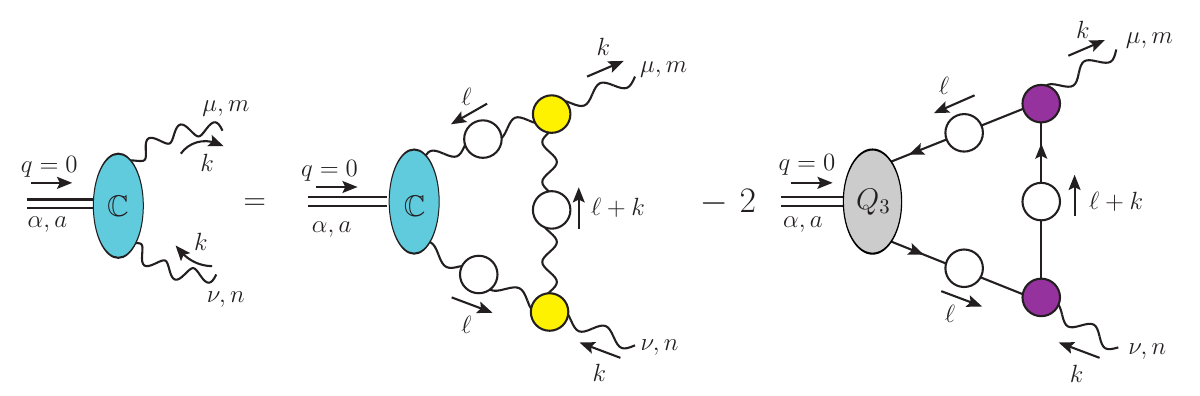}\\
\includegraphics[width=0.75\linewidth]{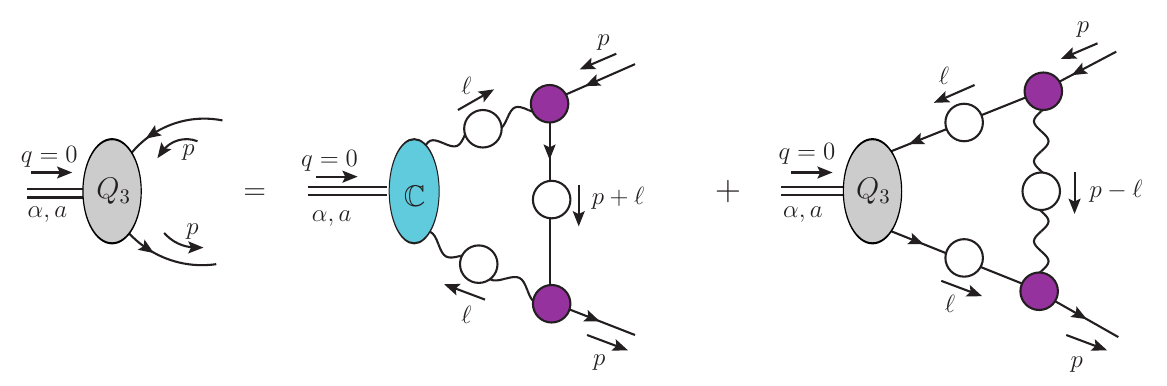}
\caption{ The system of coupled BS equations that determines the amplitudes $\Cfat(k^2)$ and $Q_3(p^2)$,  
in the one-particle 
exchange approximation.
In the first line, a summation of the quark flavors is implied, while the factor of 2 accounts for the two orientations of the quark loop.}
\label{fig:BSE}
\end{figure}

The starting point of this 
analysis is the coupled system 
of SDEs satisfied by the vertices 
$\fatg_{\alpha\mu\nu}(q,r,k)$ and 
$\fatg_\alpha(q,p_2,-p_1)$, 
shown in \fig{fig:BSE1}.
Note that we opt for the 
version of the SDE obtained within
the 3PI formalism; as a result, 
the vertices carrying the momentum 
$q$ are fully dressed. 
The corresponding four-particle 
kernels ${\cal K}_{ij}$ are appropriately 
modified to avoid overcounting; 
as may be deduced from \fig{fig:BSE}, 
for the purposes of this computation 
the ``one-particle exchange'' approximations of these kernels will 
be employed.

To obtain 
the relevant dynamical equations, 
we substitute for the vertices 
$\fatg_{\alpha\mu\nu}(q,r,k)$ and 
$\fatg_\alpha(q,p_2,-p_1)$
appearing on either side of the 
SDE system (red and orange circles)
the expressions given in \2eqs{fvert}{fvertq},
respectively. This introduces the pole parts 
on both sides of the system, and, after 
contraction of the first SDE (top row) 
by ${\cal T}_{\mu'\nu'}^{\mu\nu}(r,k)$, 
one may use 
the expressions given 
by \2eqs{Vtens}{eq:Vquark}.
Then, 
as the limit $q \to 0$ is taken on both sides of 
the SDEs, the pole terms dominate over 
the regular parts, and, after triggering 
\2eqs{eq:taylor_C}{TaylorCq},  
the displacement functions 
emerge as the leading contributions.
Finally, the proper identification 
of the various tensorial structures on both sides, and subsequent matching of their co-factors, 
gives rise to a BS system that may be 
schematically represented as 
\begin{equation}
\label{BSEcoupled}
{\cal A}_m = \sum_{n=1}^4 \int \!\! d^{\,4}\ell \,
{\cal M}_{mn} \,{\cal A}_n \quad ;\quad m=1,2,3,4\,,
\end{equation}
where the four-entry quantity
\mbox{${\cal A} = \big[\Cfat, Q_3,\Qfat_{2+3},\Qfat_1\big]$} was introduced,
and the functions 
${\cal M}_{mn}$ 
contain all remaining ingredients; 
their explicit form depends on the 
approximations employed for the kernels ${\cal K}_{ij}$.

In what follows we approximate the kernels ${\cal K}_{ij}$ by their 
{\it one-particle exchange} form, shown in \fig{fig:BSE}.
In addition, we focus exclusively on the
form factor $\lambda_1$, 
associated with the tree-level tensor $\gamma_{\alpha}$,
and determine the displacement 
function $Q_3$ appearing 
in the first relation of \1eq{SMlamdai}. 
Therefore, we restrict our 
analysis to the reduced system comprised by $\Cfat(r^2)$ and $Q_3(p^2)$ only, as shown 
in \fig{fig:BSE}. We have confirmed that the omission of the functions 
$\Qfat_{2+3}(p^2)$ and $\Qfat_1(p^2)$ has practically 
no impact on $\Cfat(r^2)$ and $Q_3(p^2)$.

In order to proceed, we need to furnish inputs for the Landau-gauge propagators and 
transversely projected vertices 
appearing in the system of \fig{fig:BSE}. 
For the gluon, ghost, and quark propagators, we use appropriate fits 
to results obtained from lattice QCD, as described in App.~\ref{app:fits}.
Furthermore, we assume that the transversely projected three-gluon and quark-gluon vertices,
\begin{align} 
\fatgb_{\alpha\mu\nu}(q,r,k) :=&\, P_{\alpha}^{\alpha^\prime}(q)P_{\mu}^{\mu^\prime}(r)P_{\nu}^{\nu^\prime}(k)\fatg_{\alpha^\prime\mu^\prime\nu^\prime}(q,r,k) \,, \nonumber\\
\fatgb_\alpha(q,p_2,-p_1) :=&\, P^{\alpha^\prime}_\alpha(q)\fatg_{\alpha^\prime}(q,p_2,-p_1) \,,
\end{align}
respectively, can be approximated by
\begin{align}
\fatgb^{\alpha\mu\nu}(q,r,k) =&\, \fatgb_{0}^{\alpha\mu\nu}(q,r,k)\Ls(s^2) \,, \qquad  s^2 = \frac{1}{2}(q^2 + r^2 + k^2) \,, \nonumber\\
\fatgb^\alpha(q,p_2,-p_1) =&\, \fatgb_{0}^\alpha(q,p_2,-p_1)\lambda_1(\bar{s}^2) \,, \qquad \bar{s}^2 = \frac{1}{2}(q^2 + p_1^2 + p_2^2) \,, \label{planar}
\end{align}
%
where $\fatgb_{0}^{\alpha\mu\nu}$ and $\fatgb_{0}^\alpha$ are the tree-level forms of $\fatgb_{\alpha\mu\nu}$ and $\fatgb_\alpha$, and are obtained by transversely projecting the $\Gamma_{\!0}^{\alpha\mu\nu}$ and $\Gamma_{\!0}^\alpha$ of Eqs.~\eqref{3g_tree} and~\eqref{qg-tree}, respectively. 
For the three-gluon vertex, \1eq{planar} has been validated by numerous studies in pure Yang-Mills theory~\mbox{\cite{Eichmann:2014xya,Blum:2014gna,Huber:2020keu,Pinto-Gomez:2022brg,Aguilar:2023qqd}}, and we assume it to hold as well for $N_f = 2$. In the case of the quark-gluon vertex, the particular form in \1eq{planar} has not been tested explicitly in the literature. However, previous works have already shown that the classical form factor of $\fatgb_\alpha(q,p_2,-p_1)$ is the largest in magnitude~\cite{Skullerud:2002ge,Skullerud:2003qu,Skullerud:2004gp,Lin:2005zd,Alkofer:2008tt,Aguilar:2014lha,Mitter:2014wpa,Cyrol:2017ewj,Gao:2021wun,Kizilersu:2021jen}, and that its angular dependence is rather weak~\cite{Kizilersu:2006et,Blum:2017uis,Blum:2015lsa,Cyrol:2017ewj}; together, these observations motivate the use of \1eq{planar} for $\fatgb^\alpha(q,p_2,-p_1)$.

In order to unify the 
description of the results, we 
rename $r^2 \to p^2$.
Next, we pass to Euclidean momenta, 
\ie \mbox{$p^2 \to -p^2_{\s E}$}, 
with $p^2_{\s E} >0$, 
introduce the 
integral measure in hyperspherical coordinates 
\be
\int \!d^{4}\!\ell_{\s E} = \frac{1}{(2\pi)^3}\int_{y,\theta} \quad ;\qquad \int_{y,\theta} := \int_0^\infty \!\!dy\, y\int_0^\pi \!\! d\theta\, s^2_\theta
\,,
\ee
and employ standard transformation properties for all Green's functions 
and form factors 
involved (see, \eg Eqs.~(5.13) to (5.16) of \cite{Aguilar:2014lha}); 
note, in particular, that 
\mbox{$\Cfat^{\s E}(p^2_{\s E}) = - \Cfat(-p^2_{\s E})$}~\cite{Aguilar:2021uwa} and 
$Q^{\s E}_3(p^2_{\s E}) = Q_3(-p^2_{\s E})$.

Defining
\begin{eqnarray}\label{sphcoor}
&& x = p^2 \,, \quad y = \ell^2 \,,\quad  p\cdot \ell = \sqrt{xy}c_\theta\,, \nonumber \\
&& z = (p + \ell)^2 = x + y + 2\sqrt{xy}c_\theta\,, \nonumber \\
&& u = (p - \ell)^2 = x + y - 2\sqrt{xy}c_\theta\,,
\end{eqnarray}
where the index ``E'' has been suppressed throughout, we arrive at the following system of coupled BS equations
\begin{align}
\Cfat(x) =&\, -\frac{\alpha_s}{\pi^2}\int_{y,\theta} \!\!f_{11}\,\Cfat(y) \,+\, \frac{\alpha_s}{6\pi^2}\int_{y,\theta} \!\!f_{12}\,Q_3(y)\,, \nonumber\\
Q_3(x) =&\, 
\frac{3\alpha_s}{4\pi^2}\int_{y,\theta} \!\!f_{21}\,\Cfat(y) \,+\, \frac{\alpha_s}{12\pi^2}\int_{y,\theta} \!\! f_{22}\,Q_3(y) \,. \label{BSE_euc}
\end{align}
The functions
$f_{ij}(x,y,\theta)$ are given by 
\begin{align}
f_{11} =&\,  \frac{s_\theta^2 c_\theta}{z}\sqrt{\frac{y}{x}}\left[ 3 ( x^2  + y^2 ) + 6 c_\theta \sqrt{xy}( x + y ) + xy( c_\theta^2 + 8 ) \right]\Delta(z)\Delta^2(y)\Ls^2(s^2) \,, \nonumber \\
f_{12} =&\, 4\sqrt{\frac{y}{x}}c_\theta\left\{3A_yB_yB_z + \big[(3 - 2s^2_\theta)y + 3\sqrt{xy}c_\theta\big]A^2_yA_z\right\} \frac{\lambda_1^2(s^2)}{(yA^2_y + B^2_y)^2(zA^2_z + B^2_z)}\nonumber\\
 &\,\hspace{0.5cm}- 6\bigg(1+\sqrt{\frac{y}{x}}c_\theta\bigg)\frac{A_z\lambda_1^2(s^2)}{(yA^2_y + B^2_y)(zA^2_z + B^2_z)} \,, \nonumber \\
%
%
f_{21} =&\, -2ys^2_\theta\bigg(1+\frac{2}{3}\sqrt{\frac{x}{y}}c_\theta\bigg)\frac{A_z}{zA^2_z + B^2_z}\,\Delta^2(y)\lambda_1^2(s^2) \,, \nonumber \\
f_{22} =&\, \left\lbrace \frac{2}{3}y\bigg[1 + \frac{2}{u}(y - \sqrt{xy}c_\theta)\bigg]\frac{s^2_\theta A^2_y}{(yA^2_y + B^2_y)} - 1 - \frac{2}{3}\frac{y}{u}s^2_\theta \right\rbrace \frac{\Delta(u)\lambda_1^2({\bar s}^2)}{(yA^2_y + B^2_y)} \,,
\end{align}
where $s^2 = x + y + \sqrt{xy}c_\theta$, ${\bar s}^2 = x + y - \sqrt{xy}c_\theta$, and the subscripts $y$ or $z$ in the functions $A$ and $B$ indicate their momentum dependence \ie $A_z :=A(z)$, etc.

\subsection{Inputs}\label{subsec:BSE_inputs}

In order to proceed with the solution of the 
system given by \1eq{BSE_euc}, we need to specify 
the form of the various components entering 
in the functions $f_{ij}(x,y,\theta)$, 
and in particular $A$, $B$, $\Delta$, $\Ls$, and 
$\lambda_1$. In addition, the ghost dressing function 
$F$ is needed for implementing the required 
conversions between renormalization schemes, as explained below. 

Note in particular the following important points: 

({\it i})
Throughout this work, we will use as external inputs a series of fits given in App.~\ref{app:fits} to the $N_f = 2$ lattice data for the gluon and ghost propagators from~\cite{Ayala:2012pb,Binosi:2016xxu}, and for $A(p^2)$, $B(p^2)$ and $\lambda_1(p^2)$ from~\cite{Oliveira:2018lln,Kizilersu:2021jen}. Specifically, for the quark functions, we employ the setups denominated ``L08'' and ``L07'' in Table I of~\cite{Kizilersu:2021jen}, which have large statistics and small current quark mass, $m_q$, and pion mass, $m_\pi$, namely \mbox{$(m_q,m_\pi) = (6.2,280)$~MeV} and \mbox{$(8,295)$~MeV}, respectively.

({\it ii})
The lattice data of~\cite{Oliveira:2018lln,Kizilersu:2021jen} 
for the functions $A(p^2)$ and $\lambda_1(p^2)$
display visible artifacts in the ultraviolet, where the known perturbative behavior of these functions is not accurately reproduced. This issue, which has been discussed in various studies~\mbox{\cite{Skullerud:2001aw,Skullerud:2004gp, Skullerud:2003qu,Lin:2005zd,Sternbeck:2017ntv,Kizilersu:2006et,Oliveira:2018lln,Kizilersu:2021jen}}, is ameliorated by the use of the so-called ``tree-level correction''~\cite{Skullerud:2001aw,Oliveira:2018lln,Kizilersu:2021jen}, but even then is not completely cured. In fact, signs of these artifacts are already present at \mbox{$p = 3$~GeV}, where the L07 $\lambda_1(p^2)$ starts to increase [see bottom left panel of \fig{fig:lambda1}], rather than continuing decreasing as predicted by perturbation theory~\cite{Pascual:1984zb,Muta:1987mz,Davydychev:2000rt}. For this reason, we discard the data for these functions for momenta above $2.5$~GeV. Since we employ fitting functions that reproduce the one-loop resumed perturbative result for each of the functions at large momenta (see App.~\ref{app:fits}), the ultraviolet behavior of our inputs is under control.

({\it iii}) 
We next turn to the 
$\Ls(p^2)$. Unfortunately, 
the only available unquenched lattice results for   
$\Ls(r^2)$ are not for $N_f = 2$
but rather for 
$N_f = 2+1$
(two light quarks with current mass $1.3$~MeV, and a heavier one, with current mass $63$~MeV)~\cite{Aguilar:2019uob}.
However, at least in the case 
of the gluon propagator~\cite{Ayala:2012pb,Aguilar:2019uob}, 
the difference between 
$N_f = 2$ to $N_f = 2+1$ is rather mild. 
It is therefore reasonable to use for the $\Ls(p^2)$ in the present analysis the $N_f = 2+1$ data from~\cite{Aguilar:2019uob}.
We hasten to emphasize that this last approximation only affects the pole amplitudes determined through the BS equations, whereas the determination of $Q_3(p^2)$, carried out in the next section through the analysis of the WI displacement, 
{\it does not} depend on $\Ls(p^2)$. 

({\it iv})
In general, the aforementioned lattice data are renormalized in different schemes; therefore, their self-consistent use hinges on their 
careful conversion into a  
common renormalization scheme. To that end, we adopt a particular variation of momentum subtraction (MOM) scheme, the so-called \MOMt{} scheme~\cite{Skullerud:2002ge}, defined by the prescription
\be 
\Delta^{-1}(\mu^2) = \mu^2\,, \quad F(\mu^2) = 1 \,, \quad A(\mu^2) = 1 \,, \quad \lambda_1(\mu^2) = 1 \,, \label{MOMtilde}
\ee
where we take $\mu = 2$~GeV. The 
procedure employed for 
consistently adjusting the results between renormalization schemes is described in detail in App.~\ref{app:renorm}.

\begin{figure}[t]
\includegraphics[width=0.45\linewidth]{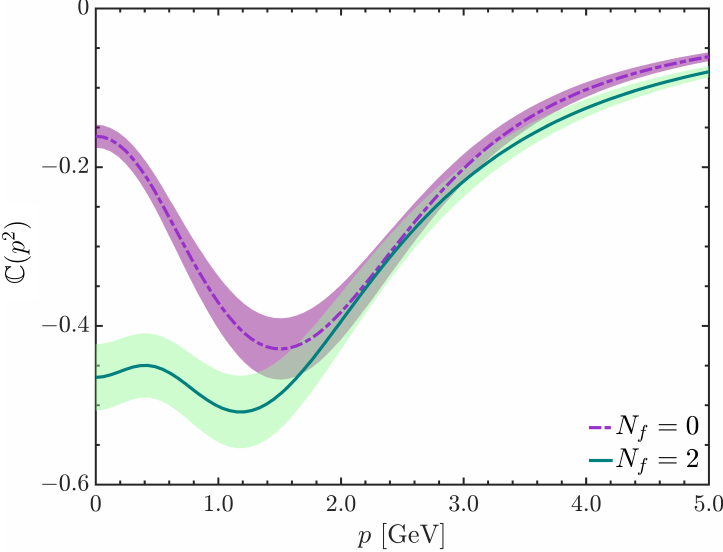}
\hfil\includegraphics[width=0.45\linewidth]{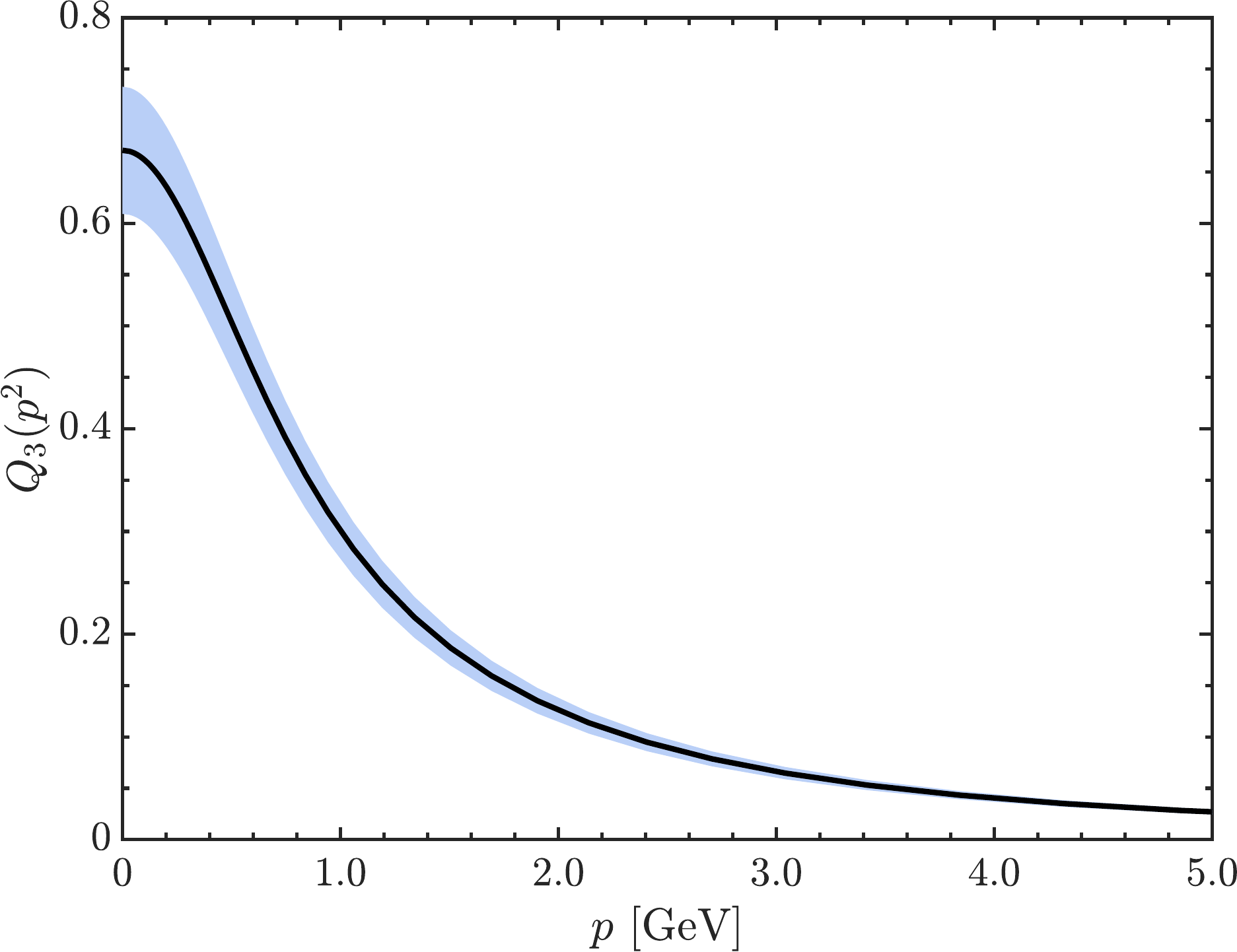}
\caption{ The 
green continuous curve on the left 
panel and the black continuous 
curve (with blue band) on the right are 
the solutions 
obtained from 
\1eq{BSE_euc} using 
unquenched ($N_f=2$) lattice 
results as inputs. 
The input used for $\lambda_1(s^2)$ is taken from the 
L08 setup of~\cite{Kizilersu:2021jen}; the results using LO7 inputs are nearly identical.
The purple dot-dashed 
curve on the left panel is the 
corresponding solution when quenched lattice 
inputs are employed (pure Yang-Mills case), and is displayed for the purpose 
of comparison.}
\label{fig:BSE_res}
\end{figure}

\subsection{Results}\label{subsec:BSE_res}

With all external ingredients consistently renormalized in the \MOMt{} scheme, we are in position to solve the BS equations of \1eq{BSE_euc} numerically, converting it to an eigenvalue problem. 
There are two important points to emphasize. 
First, as happened in the quenched case~\cite{Aguilar:2021uwa}, the coupling strength required for obtaining a solution exceeds 
the \MOMt{} value; specifically, we have 
that $\alpha^{\rm MOM}_s = 0.47$ [see \1eq{alpha_MOMt}], while the required value 
for solving the system is $\alpha^{\rm BS}_s = 1.17$\,.
The eigenvalue is known to depend 
strongly on the details of the 
truncation used for the kernels ${\cal K}_{ij}$,
but affects only slightly the form of the 
solutions obtained. 
Second,  
since the system of BS equations in \1eq{BSE_euc} is linear and homogeneous, its overall scale is undetermined: the multiplication of a given solution by an arbitrary constant 
yields another solution. 

In \fig{fig:BSE_res} we show the resulting $\Cfat(p^2)$ (left) and $Q_3(p^2)$ (right) as blue and black curves, respectively. 
This particular solution has its scale set by fitting an arbitrary solution of the BS equations to the result of the WI displacement, to be described in the next section. The bands around the curves result from the standard error in determining this scale. 

As we can see in \fig{fig:BSE_res}, for $p<1$~GeV the $\Cfat(p^2)$ and $Q_3(p^2)$ have similar magnitudes, (but opposite signs), suggesting that both amplitudes might have similar weights in the BS dynamics.  However, the terms containing $\Cfat(p^2)$ in \1eq{BSE_euc} have much larger prefactors. Moreover, we can see from \fig{fig:BSE_res} that $\Cfat(p^2)$ has support over a much longer interval of momenta than $Q_3(p^2)$, which decreases rapidly for $p>1$~GeV. As a result, $\Cfat(p^2)$ dominates the dynamics of the BS system. In fact, 
the present solution for $\Cfat(p^2)$ is rather similar to quenched results obtained previously~\cite{Aguilar:2011xe,Aguilar:2017dco,Aguilar:2021uwa}, shown 
as the purple dot-dashed curve in the left panel of \fig{fig:BSE_res}.
Finally, we point out that if $\Cfat(p^2)$ is set to zero in \1eq{BSE_euc}, the resulting BS equation for $Q_3(p^2)$ has no solutions 
for $\alpha_s > 0$.

\begin{figure}[t]
\includegraphics[width=0.49\linewidth]{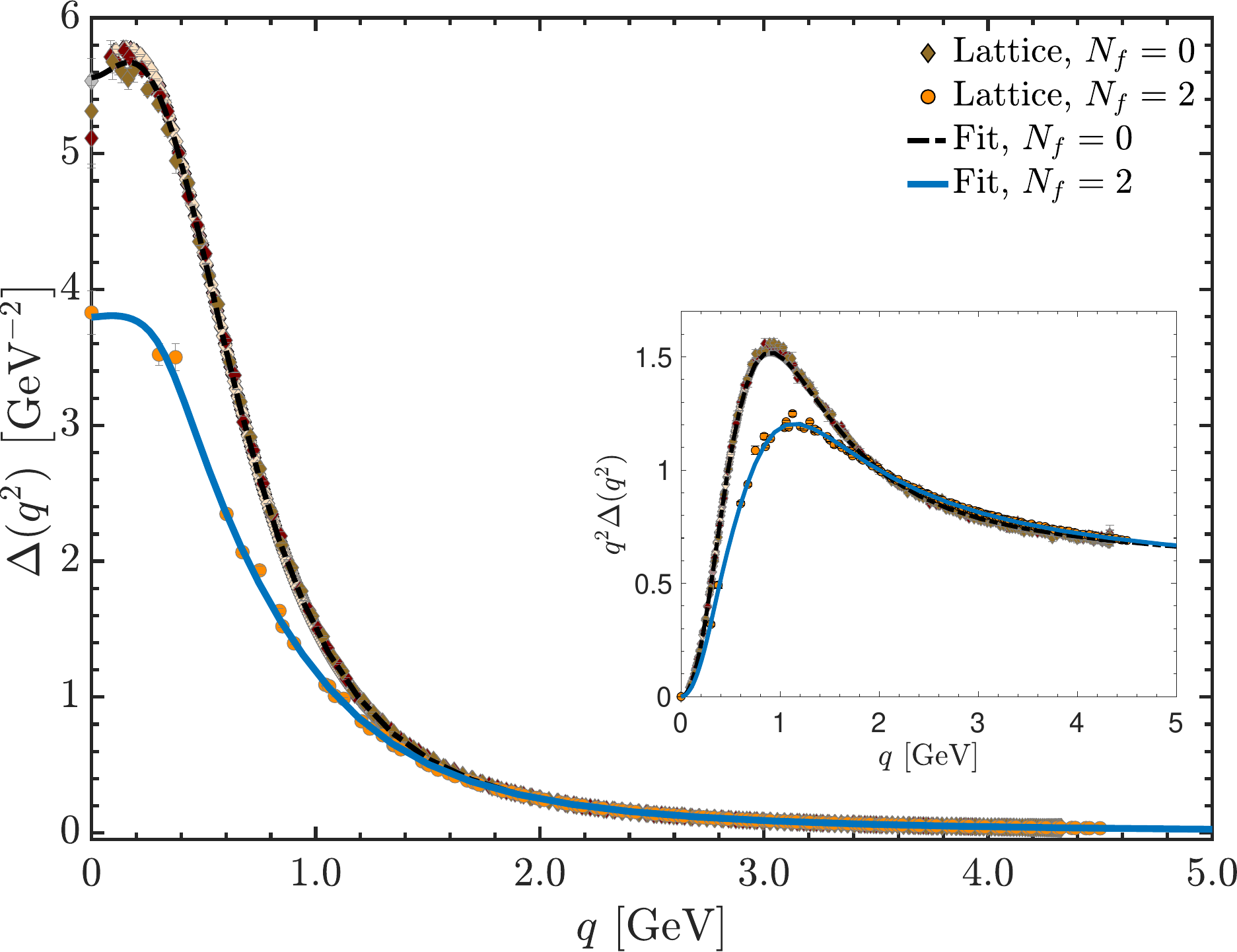}
\hfil\includegraphics[width=0.49\linewidth]{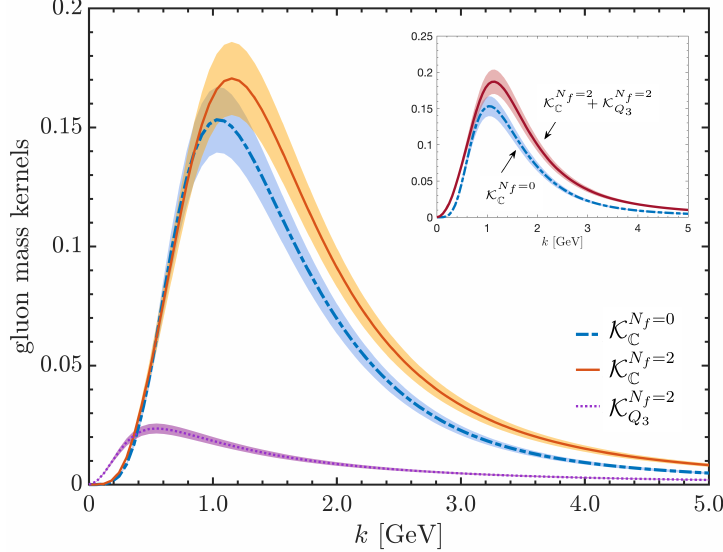}
\caption{Left panel: 
Lattice data for the gluon propagator 
for $N_f=0$ (brown diamonds)~\cite{Aguilar:2021okw}
and $N_f = 2$ (orange circles)~\cite{Ayala:2012pb,Binosi:2016xxu} .
The black dot-dashed and blue continuous curves represent the 
fits given in Eq.~(C11) of \cite{Aguilar:2021uwa} and \1eq{L_fit}, respectively.
In the inset we show the corresponding 
gluon dressing functions.
Right panel: 
The kernels 
${\cal K}_\Cfat^{N_f=0}(k^2)$ (blue dot-dashed), 
${\cal K}_\Cfat^{N_f=2}(k^2)$ (orange continuous), 
and 
${\cal K}_{Q_3}^{N_f=2}(k^2)$ (purple dotted). 
The inset shows ${\cal K}_{\Cfat}^{N_f=2}(k^2) + {\cal K}_{Q_3}^{N_f=2}(k^2)$ as a red continuous curve, compared to ${\cal K}_\Cfat^{N_f=0}(k^2)$. The error bands are 
obtained through appropriate propagation of 
the corresponding errors in the $\Cfat(p^2)$ and $Q_3(p^2)$ of \fig{fig:BSE_res}.}
\label{fig:gluon_Nf2}
\end{figure}

The solutions obtained from the BS system indicate that 
the mass generating mechanism known from the 
pure Yang-Mills case is only mildly affected by the inclusion of two active quark flavors.
To appreciate this point in some detail, 
we turn to the gluon mass equation 
that emerges from the treatment of the 
two diagrams shown in \fig{fig:pi_unqueched};
the additional gluonic corrections 
stemming from the two-loop diagrams 
will be neglected for the purposes of this qualitative analysis

The procedure outlined in~\cite{Papavassiliou:2022wrb,Ferreira:2023fva}
for obtaining the contribution to the gluon mass 
from diagram $d_1$ by appealing to 
\1eq{eq:taylor_C}
can be straightforwardly 
extended to the case of diagram $d_2$, 
where \1eq{TaylorCq}  must be employed.
Keeping only the dominant term $Q_3(p^2)$, 
one obtains (Euclidean space)
\be 
m^2 = Z_3 \int_0^\infty\!\!dy\, {\cal K}_\Cfat^{N_f}(y) \, + \, Z_2 \int_0^\infty\!\!dy\, {\cal K}_{Q_3}^{N_f}(y)  \,, \label{mass}
\ee
where the renormalization constants 
$Z_2$ and $Z_3$ are defined in \1eq{Zs_def}, 
and the kernels are given by
\be 
{\cal K}_\Cfat^{N_f}(y) = - \frac{9\alpha_s}{8\pi}y^2 \Delta^2(y) \Cfat(y) \,, \qquad {\cal K}_{Q_3}^{N_f}(y) =  \frac{\alpha_s N_f}{4\pi}\left[ \frac{ y\left( y A_y^2 + 2 B^2_y \right)Q_3(y)}{\left( y A_y^2 +  B^2_y \right)^2}\right] \,. \label{mass_kerns}
\ee
Note that the dependence of ${\cal K}_\Cfat^{N_f}(y)$ on $N_f$ is only implicit, due to the $N_f$ dependence of $\Delta(p^2)$, $\Cfat(p^2)$ and $\alpha_s$.
Instead, ${\cal K}_{Q_3}^{N_f}(y)$
depends explicitly on $N_f$, as well as implicitly.  

As is known from lattice simulation~\cite{Ayala:2012pb,Binosi:2016xxu}, 
the saturation point of the 
gluon propagator with two active quarks 
is lower with respect to the 
quenched case, as may be seen on 
the left panel of \1eq{fig:gluon_Nf2}; 
this is tantamount to saying that the gluon
mass {\it increases} when quarks are introduced.
The kernels given in \1eq{mass_kerns}
seem to capture this tendency: the 
inclusion of the quarks leads to kernels 
with additional support throughout the relevant 
region of integration. 

The origin of this relative enhancement is 
twofold. First, 
as shown on the right 
 panel of \fig{fig:gluon_Nf2}, 
${\cal K}_\Cfat^{N_f=2}(y)$ (orange continuous) 
increases slightly 
with respect to 
 ${\cal K}_\Cfat^{N_f=0}(y)$ (blue dot-dashed)  
 (11\% at the corresponding peaks).
 This increase is the net outcome of two 
 opposing tendencies: The unquenched gluon propagators suppress 
 ${\cal K}_\Cfat^{N_f=2}(y)$ with respect to 
 ${\cal K}_\Cfat^{N_f=0}(y)$, while the 
 unquenched coupling enhances it~\cite{Ayala:2012pb}. 
 Second, the kernel ${\cal K}_{Q_3}^{N_f}(y)$ (purple dotted), 
 which is absent when $N_f =0$, 
provides an additional small positive contribution, as shown on the 
right panel of \fig{fig:gluon_Nf2}.
The overall effect may be roughly appreciated 
if we set $Z_3=Z_2=1$ in \1eq{mass}
and then add the two kernels (red continuous); the end 
result, compared to the quenched case (blue dot-dashed), is shown as an inset on the right panel 
of \fig{fig:gluon_Nf2}.

This additional support of the unquenched kernel 
represents a relatively small increment 
with respect to the quenched kernel; 
therefore it is reasonable  
to expect a moderate increase in the 
gluon mass, compatible with the lattice 
findings of~\cite{Ayala:2012pb,Binosi:2016xxu}. 
The detailed calculation
of the effect requires proper renormalization 
and the inclusion in \1eq{mass} of two-loop dressed loops, not shown here; however, this task lies 
beyond the scope of the present work.

\section{Displacement 
function 
from lattice data}\label{sec:lat_disp}

In this section we will test 
the first relation in \1eq{SMlamdai}, 
which expresses the displacement 
associated with the classical  
form factor as 
\be
Q_3(p^2)=   \lambda_1^{\star}(p^2)  - 
\lambda_1(p^2) \,.
\label{Q3only} 
\ee
To that end, we use the first relation in 
\1eq{lamdai} to compute the WI prediction, $\lambda_1^{\star}(p^2)$,
and then subtract from it the lattice 
data of~\cite{Kizilersu:2021jen} for $\lambda_1(p^2)$.

The evaluation of the WI prediction is rather subtle, 
mainly due to the presence of the factor $Z_H$, the 
renormalization constant of the quark-ghost kernel.
$Z_H$ is finite in the Landau gauge; in fact, 
there exists a renormalization scheme, namely the Taylor scheme [see \1eq{Taylor}], where $Z_H = 1$. However, this scheme is \emph{not} the same as the \MOMt{} employed on the lattice determination of the Green's functions appearing in \1eq{lamdai}. Therefore, in App.~\ref{app:renorm} we determine the appropriate values for the \MOMt{} scheme with $\mu = 2$~GeV to be $Z_H = 1.120(8)$ and $Z_H = 1.121(9)$, for the L08 and L07 lattice setups, respectively.

In order to determine  
the form factors 
$K_1(p^2)$ and $K_4(p^2)$, we  
evaluate the 
one-loop dressed diagram
shown in \fig{H_and_K}, and use Eq.~\eqref{LandauH}. The full ghost-gluon vertex 
appearing in this diagram  
has the general form 
\be 
\fatg_{\!\!c}^{\mu}(r,k,q) = B_1(r^2,k^2,q^2) r^\mu + B_2(r^2,k^2,q^2) q^\mu \,, 
\label{ghost_gluon_tens}
\ee
where $r$, $k$, and $q$ stand for the momenta of the  antighost, ghost, and gluon, respectively;
at tree-level, $B_1^0 = 1$ and $B_2^0 = 0$.

Then, it is relatively straightforward to
obtain the results 
\begin{align}
K_1(x) \!=& \frac{\alpha_sZ_H}{4\pi^2}\int_{y,\theta} ( 2 + c_\theta^2)F(y)\Delta(y)B_1(y,0,y)\frac{B_z\lambda_1(s^2)}{y\left(z A^2_z + B_z^2 \right)} \,, \nonumber\\
K_4(x) \!=& -\frac{\alpha_sZ_H}{8\pi^2}\int_{y,\theta}\left( 2 + 3\sqrt{\frac{y}{x}}c_\theta + c_\theta^2 \right) \Delta(y)F(y)B_1(y,0,y)\frac{A_z\lambda_1(s^2)}{y\left(z A^2_z + B_z^2 \right)} \,, \label{Ki_SDE} 
\end{align}
where $B_1(y,0,y)$ corresponds to the 
so-called ``soft-ghost limit'' of the form factor 
$B_1$. 

The numerical evaluation of the above expressions 
proceeds through the substitution of 
$A$, $B$, $\Delta$, $\Ls$, and 
$\lambda_1$ by fits to the lattice data, 
exactly as described in Sec.~\ref{sec:bse}. For $\alpha_s$, we use the \MOMt{} values determined in \1eq{alpha_MOMt}.
Regarding $B_1(y,0,y)$, since no lattice data are available for this 
quantity for $N_f =2$, we determine it from 
the SDE that it satisfies [see App.~\ref{app:ghost-gluon}].

\begin{figure}[t]
\includegraphics[width=0.45\linewidth]{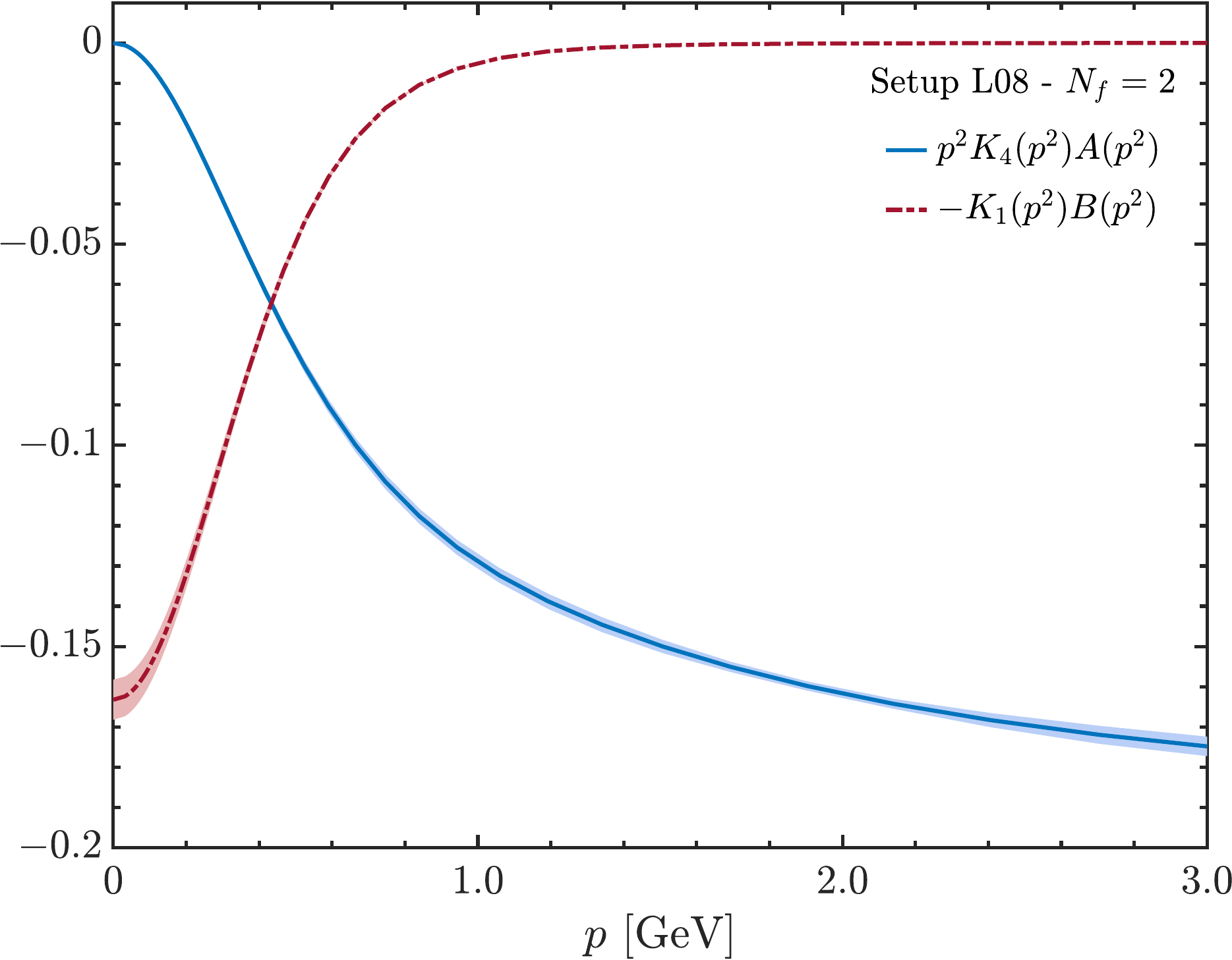}
 \hfil \includegraphics[width=0.45\linewidth]{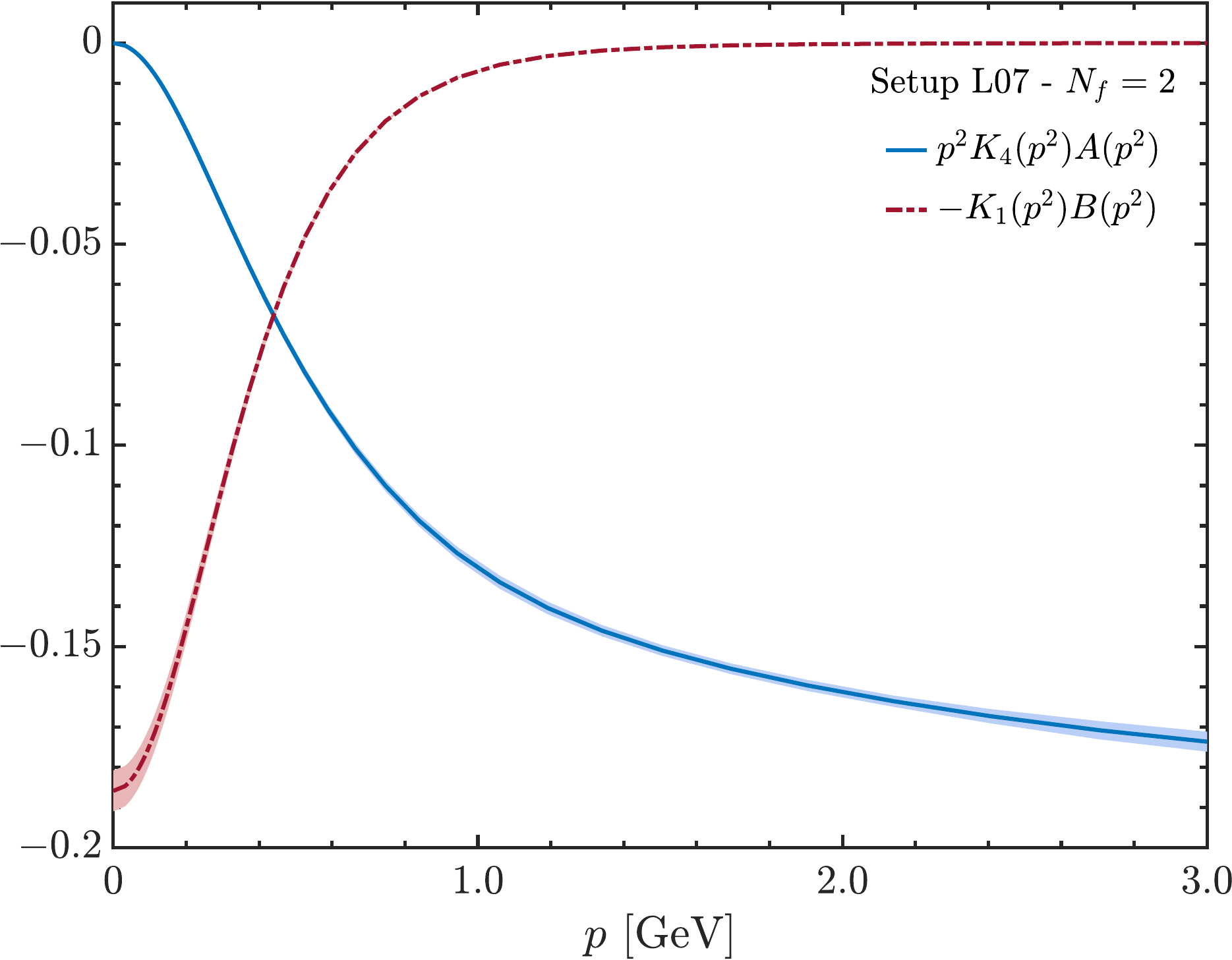}
\caption{ 
The dimensionless 
 combinations $p^2 K_4(p^2)A(p^2)$ (blue continuous) and $-K_1(p^2)B(p^2)$ (red dashed) 
 determining 
 $\lambda_1^\star(p^2)$, 
 {\it viz.} 
 first relation in
 \1eq{lamdai}. 
 The left and right panels 
 correspond to the use of 
 inputs 
 from the lattice 
 setups L08 (left) and L07 (right).}
\label{fig:QKs}
\end{figure}

The dimensionless combinations 
$p^2 K_4(p^2) A(p^2)$
and $-K_1(p^2)B(p^2)$ 
appearing in \1eq{lamdai} 
are shown 
in \fig{fig:QKs} as blue continuous and red dashed lines, respectively.
The two panels correspond to the results obtained when the inputs used for 
$\lambda_1$ 
in \1eq{Ki_SDE} originate 
from the lattice setup 
L08 (left) or L07 (right).
The comparison of the two panels 
reveals 
that $p^2 K_4(p^2)A(p^2)$ is nearly identical for both lattices, whereas the value of $-K_1(p^2)B(p^2)$ at the origin is slightly larger in magnitude when the 
$\lambda_1$ from the  
L07 setup is used.

\begin{figure}[t]
\includegraphics[width=0.45\linewidth]{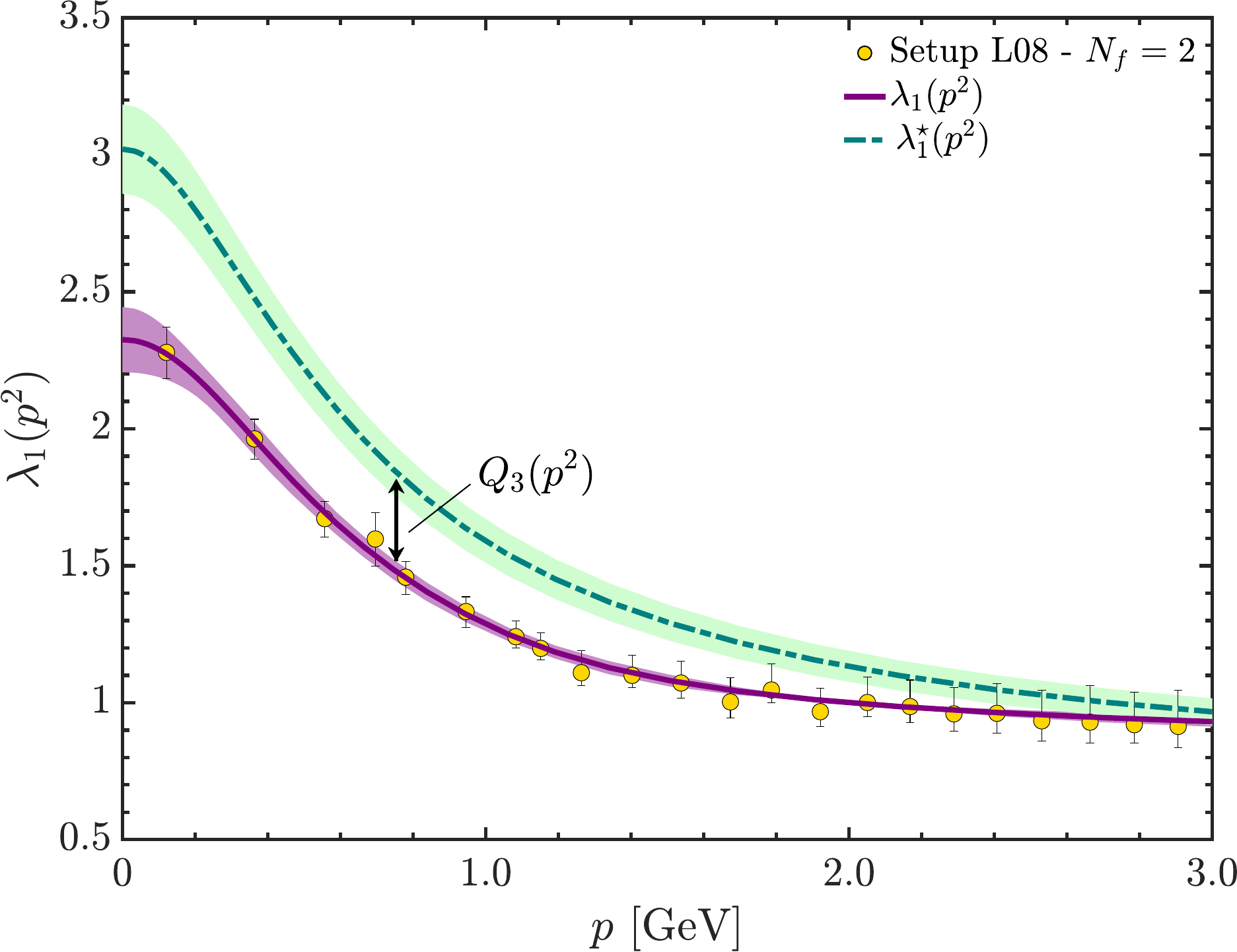} \hfil 
\includegraphics[width=0.45\linewidth]{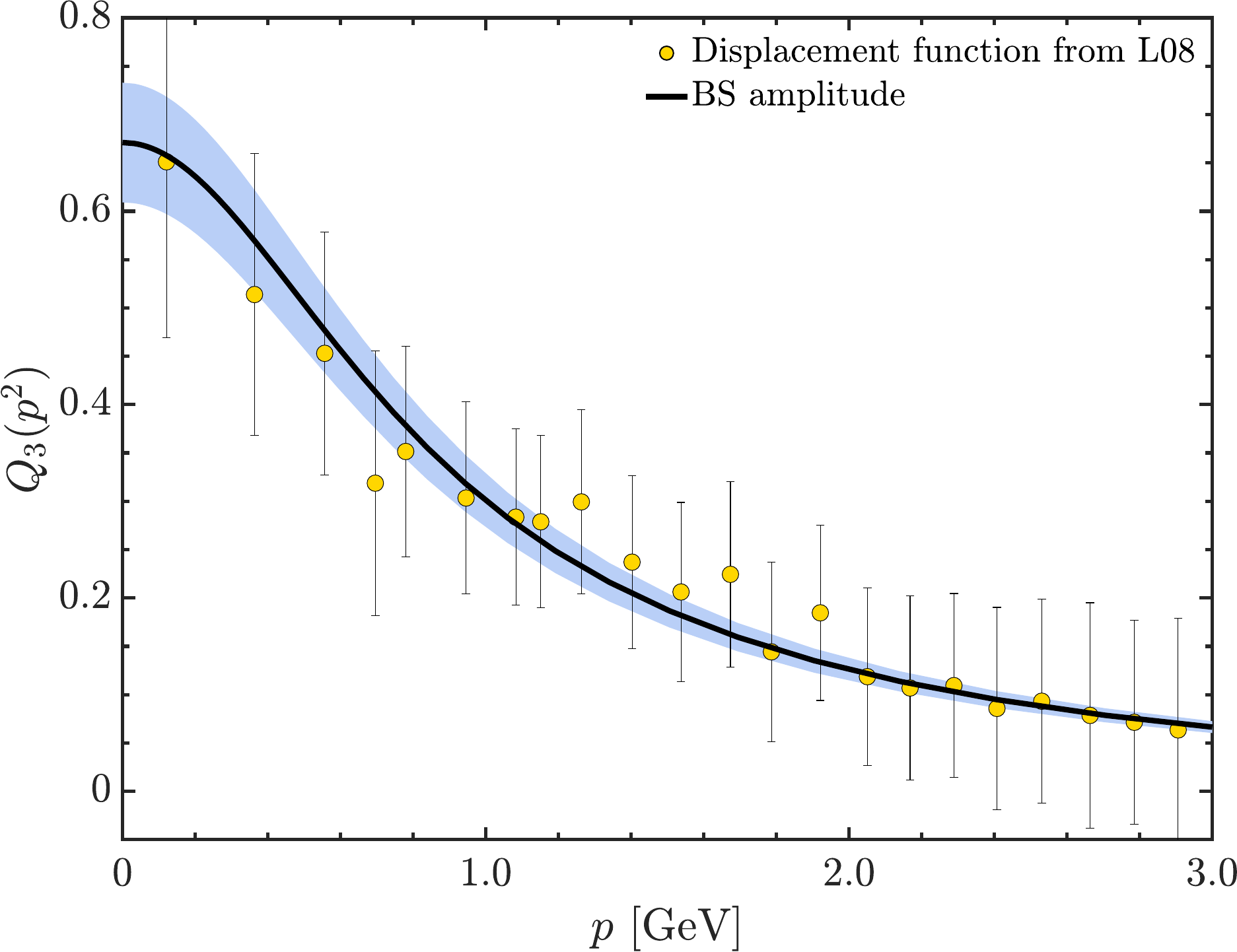}\\
\includegraphics[width=0.45\linewidth]{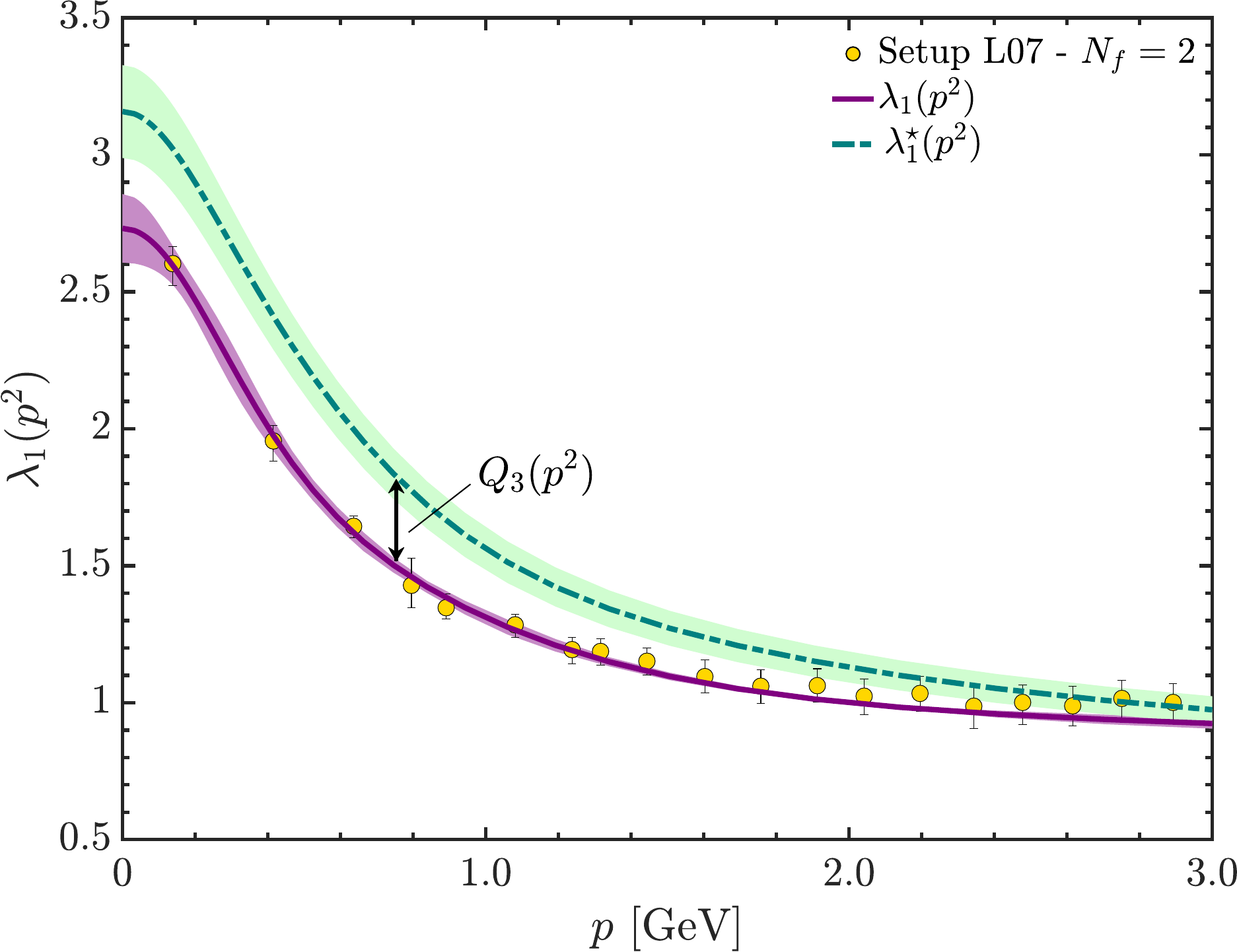} \hfil 
\includegraphics[width=0.45\linewidth]{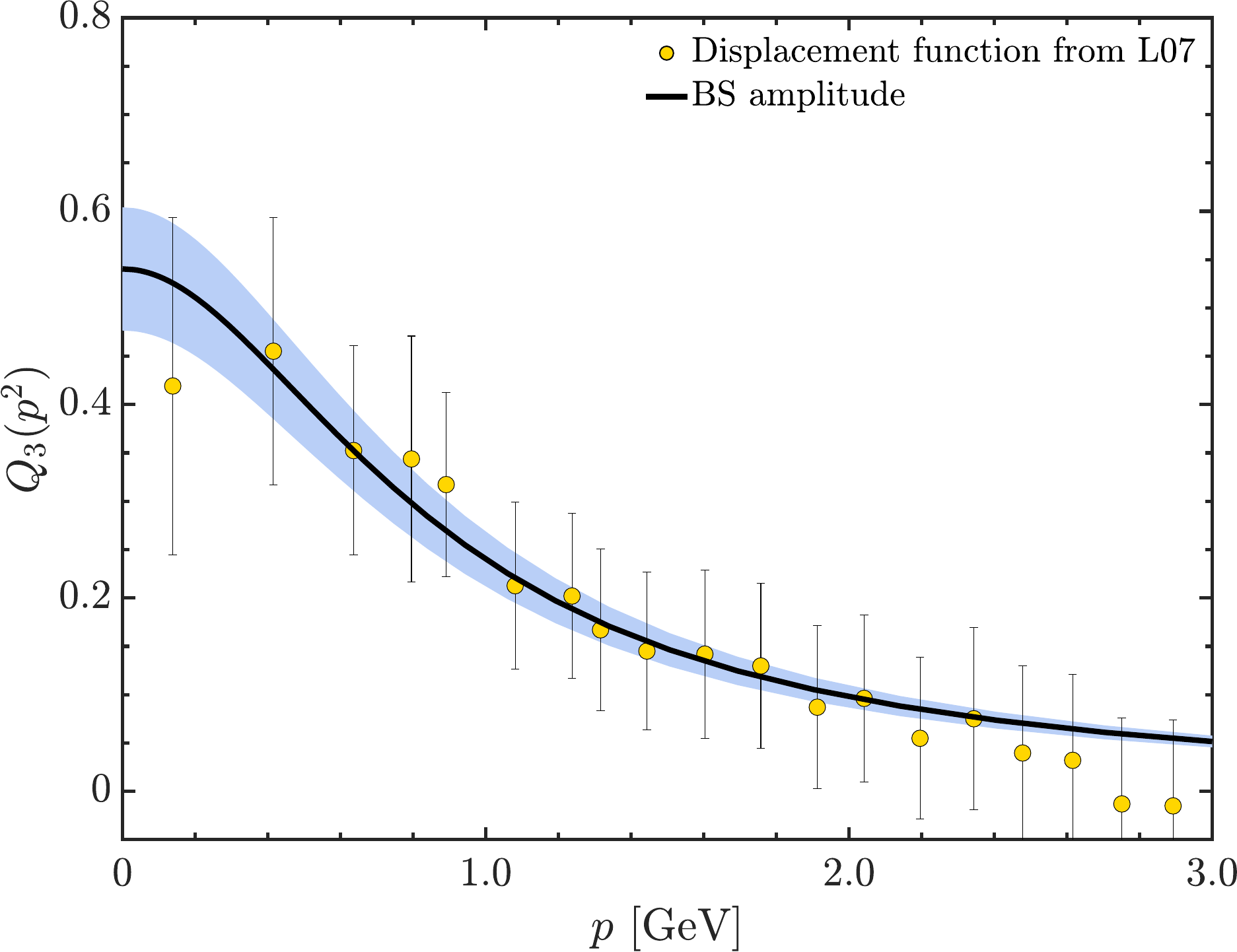}
\caption{ Top left: 
({\it i}) the L08 lattice data (points) for 
$\lambda_1(p^2)$~\cite{Kizilersu:2021jen}, 
the fit 
of \1eq{lambda_1_fit} (purple continuous),
and the 
$1\sigma$ confidence band; 
({\it ii})
the WI prediction $\lambda_1^\star(p^2)$ (green dot-dashed line and band). 
Top right: The 
displacement function 
$Q_3(p^2)$ (points) obtained as the difference
between the two curves 
shown on the left, 
as dictated by \1eq{Q3only}.
The black solid line and blue band indicate the
prediction of the 
BS equation. 
The bottom panels show the same quantities but for the L07 setup.} 
\label{fig:lambda1}
\end{figure}

Evidently,  
the errors associated with the 
inputs used to 
evaluate \1eq{Ki_SDE}
get propagated to 
the resulting 
$K_i(p^2)$. 
It turns out that 
the largest source of uncertainty 
is the form factor
$\lambda_1$.
In order 
to estimate the error that 
$\lambda_1$ introduces 
to the $K_i(p^2)$, we repeat the numerical evaluation of \1eq{Ki_SDE} with
\be 
\lambda_1(p^2) \to \lambda_1(p^2) \pm \delta \lambda_1(p^2) \,,
\ee
where $\delta \lambda_1(p^2)$ is the $1\sigma$ spread in $\lambda_1(p^2)$, shown as green bands in the left panels of \fig{fig:lambda1}, for each of the lattice setups. 
Then, the error obtained 
is combined in quadrature 
with all other intrinsic errors stemming from the remaining ingredients, mainly 
$A(p^2)$ and $B(p^2)$, 
to furnish our 
estimate for the total error 
of the quantities $p^2 K_4(p^2)A(p^2)$ and $-K_1(p^2)B(p^2)$.

With the $K_i(p^2)$ in hand, we use the first line of \1eq{lamdai} to compute $\lambda_1^\star(p^2)$. The result is shown as a green dot-dashed line for each of the lattice setups, L08 and L07, on the top left and bottom left panels of \fig{fig:lambda1}, respectively. The green bands around these curves are our error estimates, which combine in quadrature the errors in all of the ingredients appearing in \1eq{lamdai}. In particular, we included in the total error budget a $5\%$ error estimate in the value of $F(0) = 2.39$ obtained by extrapolating the lattice data of~\cite{Ayala:2012pb,Binosi:2016xxu} 
to the origin, $p=0$.

As is clear from the left panels of \fig{fig:lambda1}, our results for $\lambda_1^\star(p^2)$ are considerably larger than the input $\lambda_1(p^2)$, indicating the existence of a nontrivial displacement, $Q_3(p^2)$. On the right panels of \fig{fig:lambda1} we show as points the results for $Q_3(p^2)$ obtained from \1eq{SMlamdai}, for the L08 and L07 setups on the top and bottom, respectively. The error bars in the points for $Q_3(p^2)$ result from combining in quadrature the standard error of the lattice $\lambda_1(p^2)$ and the estimated error in $\lambda_1^\star(p^2)$.

We next evaluate the statistical significance of the signal obtained for $Q_3(p^2)$, in comparison to the ``null hypothesis'', 
namely the case when 
$Q_3 \to Q_3^0 = 0$, due to the 
the absence of the Schwinger mechanism. 
To that end, we compute separately for each of the lattice setups the $\chi^2$ deviation from the null hypothesis as
\be 
\chi^2 = \sum_{i = 1}^{ n_p } \frac{[ Q_3(p_i^2) - Q_3^0(p_i^2) ]^2}{\epsilon_{Q_3(p_i^2)}^2}\,,
\label{error}
\ee
where the sum over the 
lattice points $i$, 
corresponding to momenta $p_i$, extends until \mbox{$p_i < 2.5$~GeV};  
the reason for implementing this 
cut is explained in item ({\it ii}) of 
Sec.~\ref{subsec:BSE_inputs}.
The number of lattice points in this interval is \mbox{$n_p =18$} for the setup L08 and \mbox{$n_p =16$} for the setup L07. 
The quantity $\epsilon_{Q_3(p_i^2)}$ represents the error in the corresponding value of $Q_3(p_i^2)$. Then, 
the use of \1eq{error}
yields \mbox{$\chi^2= 119$} for the setup L08, and \mbox{$\chi^2= 76$} for the setup L07.

At this point, we compute the probability, $P_{Q_3^0}$, that our result for $Q_3$ is consistent with the null hypothesis through
\be 
P_{Q_3^0} = \int_{\chi^2}^\infty \chi^2_{\srm{PDF}}(n_p,x) dx = \frac{\Gamma(n_p/2,\chi^2/2)}{\Gamma(n_p/2)} \,,
\ee
where $\chi^2_{\srm{PDF}}$ is the $\chi^2$ probability distribution function. Then, using the above quoted values of $n_p$ and $\chi^2$ we obtain \mbox{$P_{Q_3^0} = 6.48\times 10^{-17}$} for L08, and \mbox{$P_{Q_3^0} = 8.68\times 10^{-10}$} for L07. In terms of confidence levels to discard the null hypothesis, these probabilities translate to $8.4\sigma$ and $6.1\sigma$, respectively.

\section{Discussion and Conclusions}\label{sec:conc}

The non-Abelian implementation of the SM
is associated with the appearance of longitudinally 
coupled massless poles in the fundamental vertices of the theory.
In addition to endowing the gluons with an effective mass, the action of the SM leads 
to the displacement of  
the Ward identities satisfied by the 
fundamental vertices of the theory by amounts proportional to 
the momentum-dependent residues of these poles. 
This property is particularly crucial, because it enables us to
confirm the nonvanishing nature of the pole residues by 
forming the difference between the lattice data and the WI prediction 
for certain vertex form factors in the soft-gluon limit.  
In the present work we have focused 
on the manifestation of this characteristic effect at the level of the quark-gluon vertex of QCD with two light quark flavors, 
$N_f=2$. 

Our analysis demonstrates that 
the quark-gluon vertex contains 
massless poles: they correspond to
the propagators of 
scalar, color-carrying 
bound states, 
which are 
composed out
of two gluons or of a quark-antiquark pair.
The residue functions associated 
with these poles play the role of the BS amplitudes 
for the bound-state formation, 
and are determined from a coupled system of BS equations.
Crucially, this system 
involves also the unquenched version of 
the function $\Cfat(p^2)$, known from 
the  well-studied case 
of the (quenched) three-gluon vertex.

The residue function $Q_3(p^2)$, associated with the 
tree-level Dirac structure $\gamma_{\alpha}$,
plays a prominent role in this study, 
because it is directly  
responsible for the displacement  
produced between the lattice data 
and the WI prediction for the classical form factor
of the quark-gluon 
vertex.  In fact, the 
detailed comparison with the 
lattice data of~\cite{Kizilersu:2021jen}
reveals a statistically significant 
discrepancy (signal), 
whose momentum-dependence is completely consistent 
with the BS solution for $Q_3(p^2)$,
as may be seen in 
\fig{fig:lambda1}.
This finding further 
corroborates the 
action of the SM 
in the context of 
unquenched QCD, providing additional evidence for the 
robustness of this 
particular mass generating scenario. 

The robustness of the result
shown in \fig{fig:lambda1}
hinges on one's ability to accurately determine the WI prediction for the form 
factor $\lambda_1(p^2)$, 
namely the $\lambda_i^{\star}(p^2)$
given by \1eq{lamdai}. 
There, in addition to the 
components $A(p^2)$ and $B(p^2)$ 
of the quark propagator, the 
form factors $K_1(p^2)$ and 
$K_4(p^2)$, associated with the 
quark-ghost kernel 
[see \1eq{LandauH} and \fig{H_and_K}],
play a central role. 
The determination of 
$K_1(p^2)$ and 
$K_4(p^2)$ involves the 
fully-dressed quark-gluon vertex, 
which has been approximated by the 
Ansatz given in the second line of \1eq{planar}.
This particular form is 
inspired by the special property 
of ``planar degeneracy'', satisfied at a high level of 
accuracy by the three-gluon vertex~\cite{Huber:2020keu,Eichmann:2014xya,Blum:2014gna,Pinto-Gomez:2022brg,Aguilar:2023qqd}.  
In the case of the quark-gluon vertex,  
the veracity of this Ansatz has not been confirmed at
a corresponding level of accuracy, even though 
it seems fairly plausible, given the 
observations made below \1eq{planar}. To be sure,
a more detailed analysis of this entire issue 
is required, in order to determine the 
limitations of this approximation, and the possible 
modifications that may arise in the form factors   
$K_1(p^2)$ and 
$K_4(p^2)$. We hope to embark on such a study in the near future. 

Finally, even though we have focused on the 
displacement 
produced to $\lambda_1(p^2)$ by 
$Q_3(p^2)$, the form factors 
$\lambda_2(p^2)$ and $\lambda_3(p^2)$
get also displaced by 
the functions $\Qfat_{2+3}(p^2)$ and $\Qfat_1(p^2)$, 
respectively
[{\it viz.} second and third relation in \1eq{SMlamdai}].
In fact, a preliminary analysis 
indicates that in the case 
of $\lambda_3(p^2)$ a detectable displacement may be obtained. 
This happens despite the fact that 
$\Qfat_1(p^2)$ is 
significantly 
smaller compared to 
$\Cfat(r^2)$ and $Q_3(p^2)$, 
because $\lambda_3(p^2)$  is 
significantly smaller to 
$\lambda_1(p^2)$, and the available  lattice 
data display rather small error bars. It is clearly important to 
explore this possibility further, 
establishing the statistical 
significance of the resulting displacement and its similarity to 
the BS prediction.

\section{Acknowledgments}
\label{sec:acknowledgments}

We thank J. Rodr{\'i}guez-Quintero and O. Oliveira for useful communications.
The work of  A.~C.~A. is supported by the CNPq grant \mbox{307854/2019-1}. A.~C.~A also acknowledges financial support from project 464898/2014-5 (INCT-FNA). D.~I. acknowledges the support provided by the University Centre EDEM. M.~N.~F. and J.~P. are supported by the Spanish MICINN grant PID2020-113334GB-I00. M.~N.~F. acknowledges financial support from Generalitat Valenciana through contract \mbox{CIAPOS/2021/74}. J.~P. also acknowledges funding from the Generalitat Valenciana grant CIPROM/2022/66.


\appendix

\section{Input fits}\label{app:fits}

In this Appendix we present the fits to lattice data for the
propagators and vertex form factors 
used as external inputs in our calculations. These fits capture a variety of physical properties in the infrared, discussed elsewhere~\cite{Aguilar:2019uob,Aguilar:2021lke,Ferreira:2023fva}, and reproduce the one-loop anomalous dimensions of the corresponding functions in the ultraviolet~\cite{Davydychev:1996pb,Davydychev:2000rt}.

The fitting functions employed 
share certain basic building blocks, which differ only in the specific values 
that some sets of parameters acquire 
in each case. In particular, 
the parameters $\delta$, $\kappa$, $\eta_i$, and $b_i$ appear in various fits, taking distinct numerical 
values in each case. The sets of numerical values employed are displayed in 
Table~\ref{tab:pars}.

\begin{table}[t]
\begin{tabular}{|c|c|c|c|c|c|c|c|c|}
\hline
  & $\delta$ & $\kappa^2$~[GeV$^2$] & $\eta_1^2$~[GeV$^2$] & $\eta_2^2$~[GeV$^2$] & $b_0$ & $b_1$~[GeV$^2$] & $b_2$~[GeV$^2$]& $b_3$~[GeV$^2$]  \\
\hline
$\Delta(r^2)$  & $0.112$ & $71.8$ & $10.1$ & $0.895$ & $-0.0998$ & $-1.67$ & $0.684$ & $0.321$ \\
\hline
$F(r^2)$  & -- & -- & $2.68$ & $1.27$ & $-0.440$ & $6.23$ & $0.820$ & $23.2$\\
\hline
$\Ls(r^2)$  & $0.0629$ & $12.3$ & $1.00$ & $1.48$ & $0.102$ & $25.9$ & $1.70$ & $19.0$ \\
\hline
$A(p^2)$ [L08]  & -- & $0.930$ & -- & -- & $0.360$ & $0.642$ & $0.175$ & $0.462$  \\
\hline
$A(p^2)$ [L07]  & -- &  $2.69$ & -- & -- & $0.849$ & $0.351$ & $0.193$ & $1.98$ \\
\hline
${\cal M}(p^2)$ [L08]  & $0.294$ & $0.520$ & $1.31$ & $13.0$ & -- & -- & -- & --  \\
\hline
${\cal M}(p^2)$ [L07]  & $0.110$ & $0.596$ & $1.00$ & $31.1$ & -- & -- & -- & -- \\
\hline
$\lambda_1(p^2)$ [L08]  & -- & -- & $24.6$ & $0.0371$ & $-0.587$ & $11.8$ & $0.599$ & $1.03$ \\
\hline
$\lambda_1(p^2)$ [L07]  & -- & -- & $24.6$ & $0.0371$ & $-0.483$ & $2.33$ & $0.495$ & $0.745$ \\
\hline
\end{tabular}
\caption{ Fitting parameters used for Eqs.~\eqref{L_fit}, \eqref{A_fit}, \eqref{M_fit}, and \eqref{lambda_1_fit}. In addition, for ${\cal M}(p^2)$ we use \mbox{$m_q = 6.2$~MeV}, and \mbox{$8$~MeV} for the setups L08 and L07, respectively, and \mbox{$m_0 = 345$~MeV} for both. Lastly, we employ \mbox{$\Lambda = 610$~MeV} for all quantities.}\label{tab:pars}
\end{table}

For the unquenched quantities  $\Delta(r^2)$, $F(r^2)$ and $\Ls(r^2)$, we employ the functional forms
\begin{align} 
\Delta^{-1}(r^2) =&\, r^2 \left[ \frac{\delta}{ 1 + ( r^2/\kappa^2) }\ln\left( \frac{r^2}{\mu^2} \right) + \uf^{d_{\s A}}(r^2) \right] + \nu^2 R(r^2) \,, \quad & d_{\s A} =&\, \frac{39 - 4N_f}{6\beta_0} \,, \nonumber\\ 
F^{-1}(r^2) =&\, \uf^{d_{\s c}}(r^2) + R(r^2) \,, \quad & d_{\s c} =&\, \frac{9}{4\beta_0} \,, \nonumber\\
\Ls(r^2) =&\,  1.16\left\lbrace \frac{\delta}{ 1 + ( r^2/\kappa^2) }\ln\left( \frac{r^2}{\mu^2} \right) + \uf^{d_{\s{3g}}}(r^2) + R(r^2) \right\rbrace \,, \quad & d_{\s{3g}} =&\, \frac{51 - 8N_f}{12\beta_0} \,, 
\label{L_fit}
\end{align}
where we define $\beta_0 = 11 - 2N_f/3$, 
\be 
\uf(r^2) := 1 + \frac{\ln\left[  ( r^2 + \eta^2(r^2) )/( \mu^2 + \eta^2(r^2)) \right]}{\ln\left(\mu^2/\Lambda^2\right)} \,, \qquad \eta^2(r^2) := \frac{\eta_1^2}{1 + r^2/\eta_2^2} \,,
\label{UVlog}
\ee
\be
\label{rational} 
R(r^2) :=\frac{ b_{0} + r^2/b_{1}^2 }{1 + r^2/ b_{2}^2  +  (r^2/b_{3}^2)^2 } - \frac{ b_{0} + \mu^2/b_{1}^2  }{1 + \mu^2/ b_{2}^2  +  (r^2/b_{3}^2)^2 } \,,
\ee
and the parameter \mbox{$\nu = 1$~GeV} makes the dimensionalities of $R(r^2)$ and $\Delta^{-1}(r^2)$ consistent without changing the dimensions of the parameters $b_i$. Note that the factor of $1.16$ in $\Ls(r^2)$ converts the lattice data  of~\cite{Aguilar:2019uob} to the \MOMt{} scheme [see \1eq{Z3conv}].

The lattice data of~\cite{Oliveira:2018lln,Kizilersu:2021jen} for the quark propagator are given in terms of the quark wave function, $A^{-1}(p^2)$, and running mass, ${\cal M}(p^2) := B(p^2)/A(p^2)$. These are fitted with
\be
A(p^2) = \frac{T(p^2)}{T(\mu^2)} \,, \qquad T(p^2) := \frac{ b_0 + p^2/b_1 + ( p^2/\kappa^2 )^2 }{ 1 + p^2/b_2 + ( p^2/b_3 )^2 } \,, \label{A_fit}
\ee
and
\be
{\cal M}(p^2) = \frac{m_0}{ 1 + ( p^2/\kappa^2 )^{ 1 + \delta } } + m_q\left[ \frac{1}{2}\ln\left(\frac{p^2 + \eta^2(p^2)}{\Lambda^2 + \eta^2(p^2)}\right)\right]^{- d_{\s{ {\cal M} }}}\,, \qquad d_{\s{ {\cal M} }} = \frac{4}{\beta_0}\,. \label{M_fit}
\ee

Finally, the fit for $\lambda_1(p^2)$ is obtained with
\be
\lambda_1(p^2) = \left[ \uf^{d_{\s{qg}}}(p^2) + R(p^2)\right]^{-1} \,, \qquad d_{\s{qg}} = \frac{9}{4\beta_0} \,. \label{lambda_1_fit}
\ee

\begin{figure}[ht]
\includegraphics[width=0.45\linewidth]{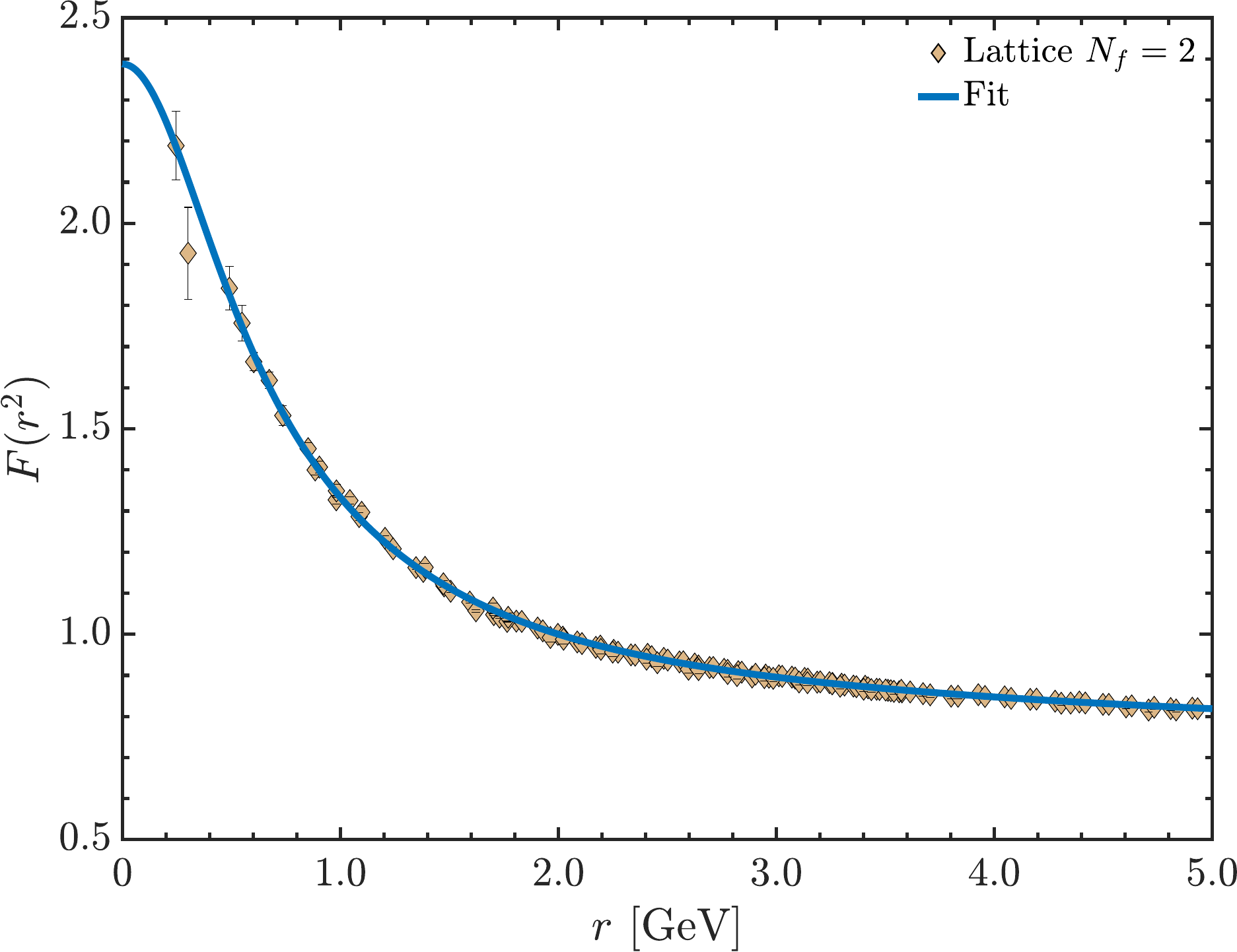} \hfil
 \includegraphics[width=0.45\linewidth]{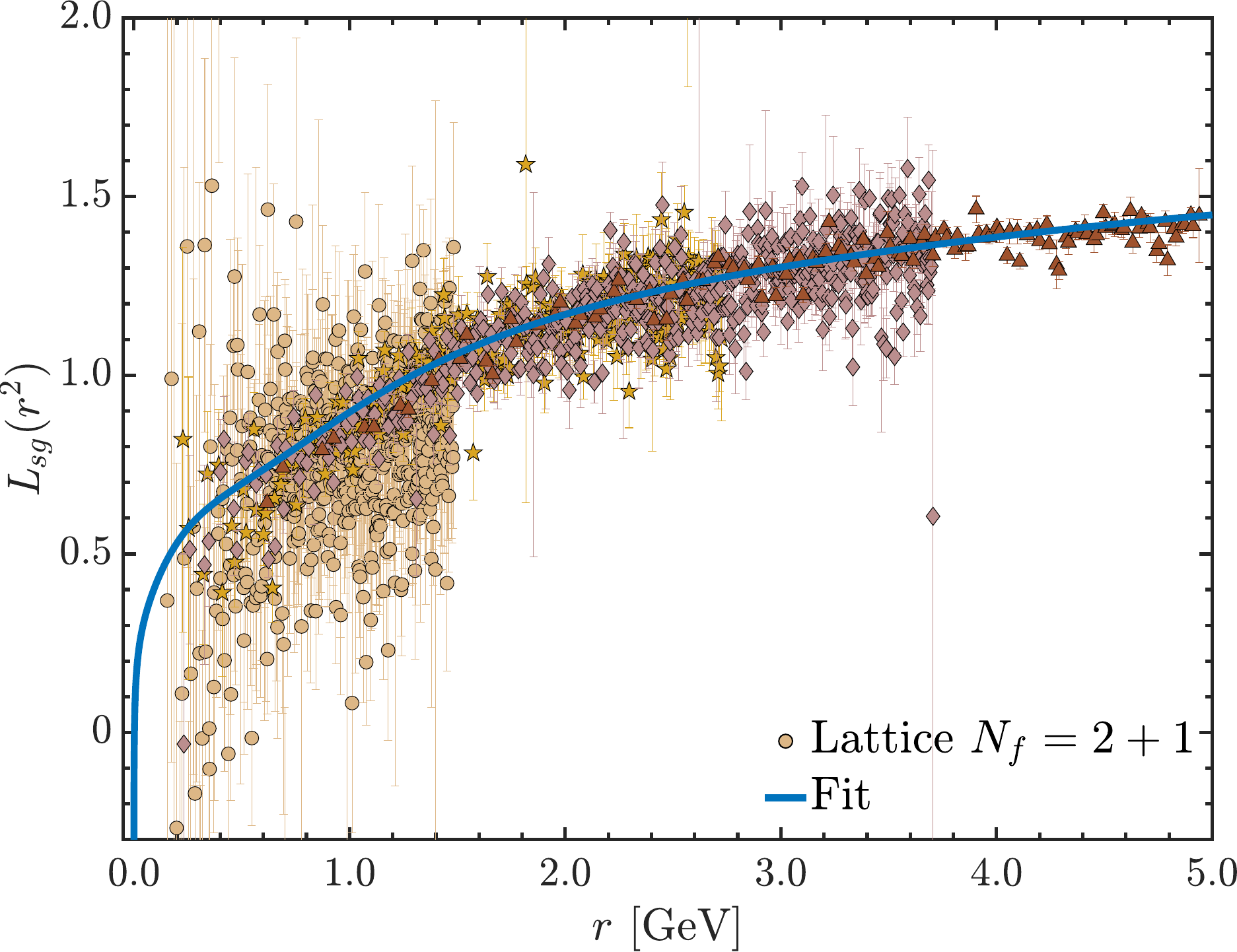}\\
\includegraphics[width=0.45\linewidth]{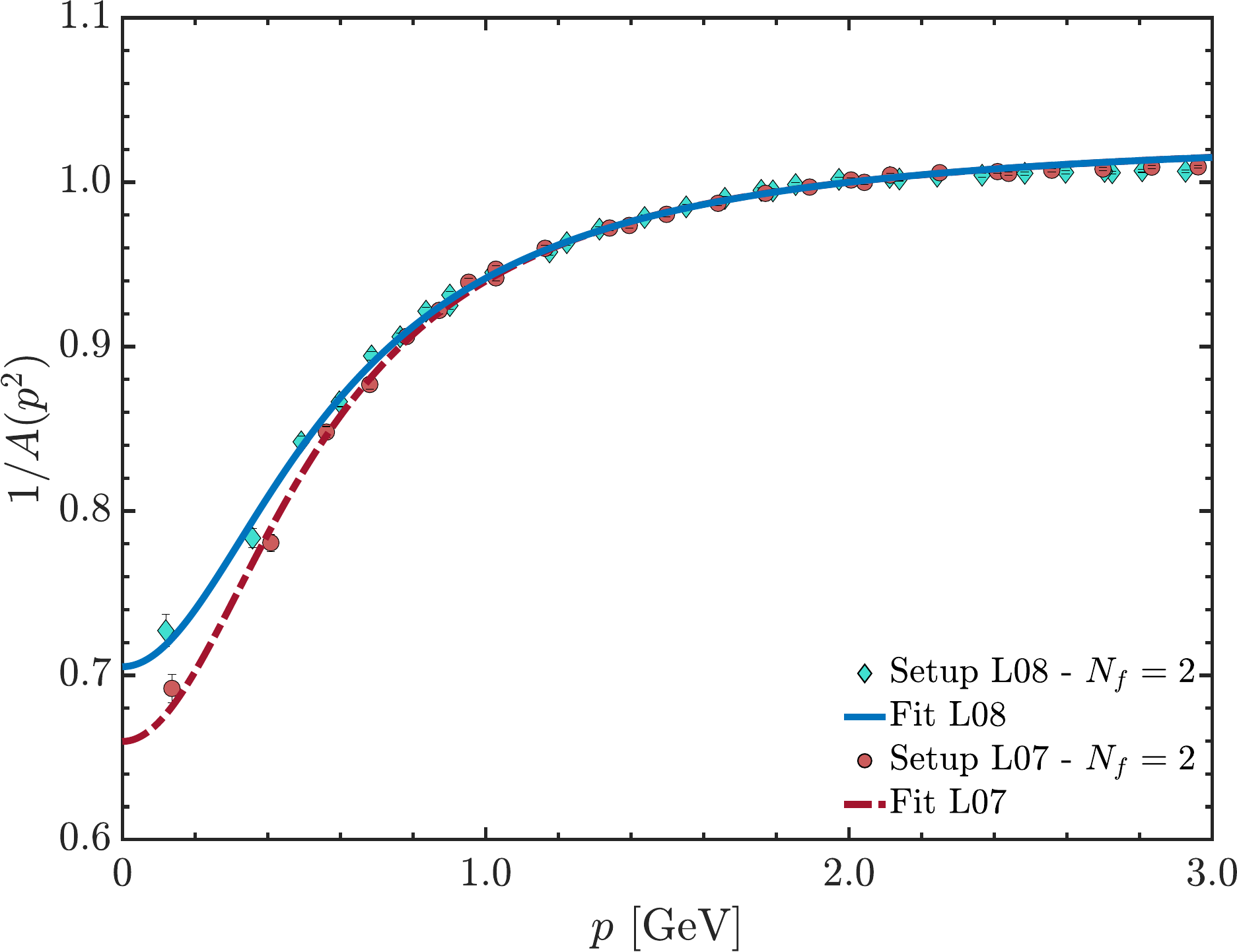} \hfil
 \includegraphics[width=0.45\linewidth]{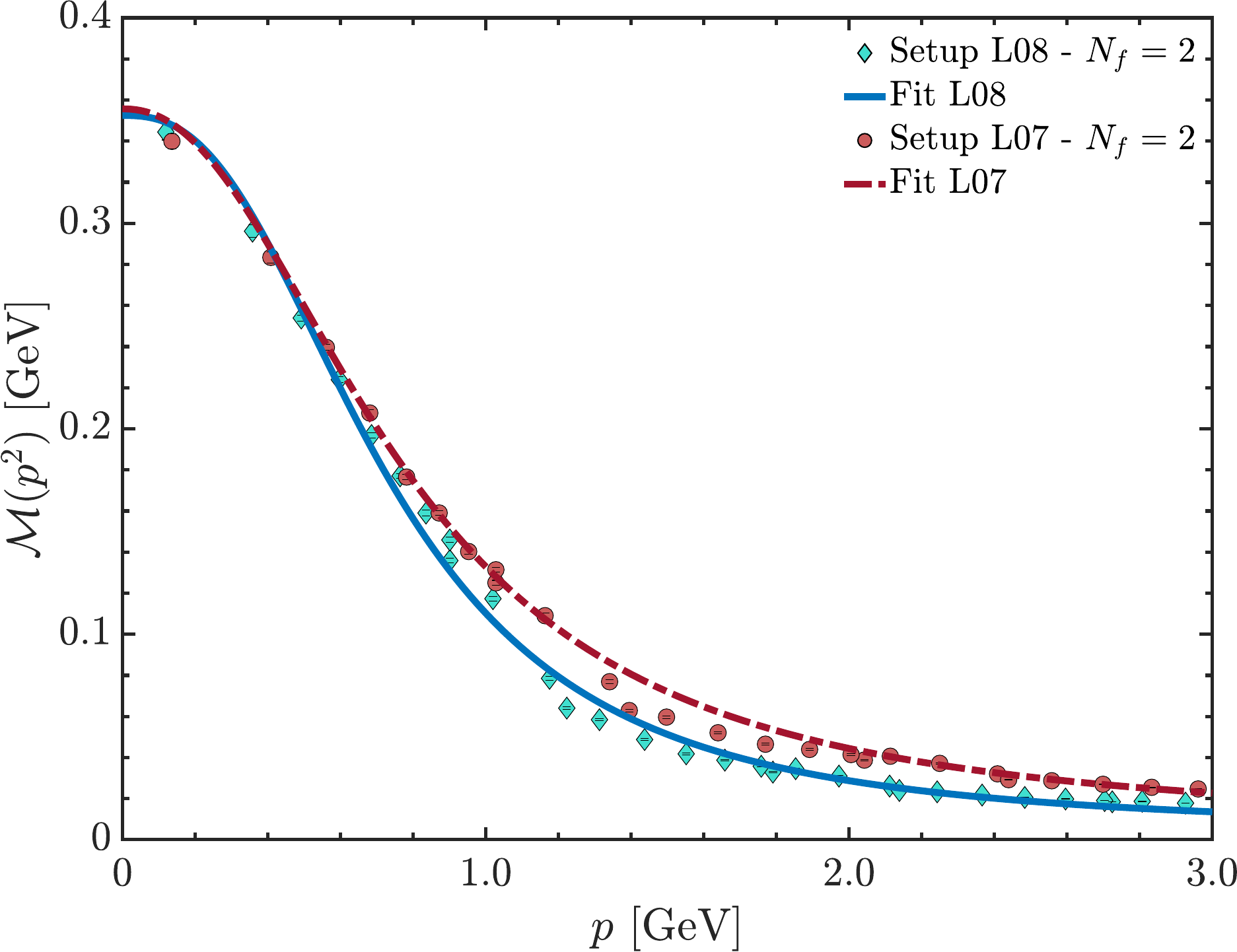}
\caption{ Top: lattice data (points) for $F(r^2)$ from~\cite{Ayala:2012pb,Binosi:2016xxu} (left) and for $\Ls(r^2)$ from~\cite{Aguilar:2019uob} (right); the 
corresponding fits 
(blue continuous) 
are given by \1eq{L_fit}. Note that for $\Ls(r^2)$ we use $N_f = 2+1$ data, as discussed in item the ({\it iii}) of Sec.~\ref{subsec:BSE_inputs}.
Bottom: lattice data of~\cite{Oliveira:2018lln,Kizilersu:2021jen} for the quark wave function (left) and running mass (right), and the fits of \2eqs{A_fit}{M_fit}. The lattice data for the setups L08 and L07 of~\cite{Oliveira:2018lln,Kizilersu:2021jen} are represented by blue and red points, respectively, and their fits by blue continuous and red dashed curves. }
\label{fig:inputs}
\end{figure}

The values of the fitting parameters are collected in Table~\ref{tab:pars} and its caption. The resulting curves for $\Delta(r^2)$ and $\lambda_1(p^2)$ are compared to the corresponding lattice data in the left panels of Figs.~\ref{fig:gluon_Nf2} and \ref{fig:lambda1}; the functions $F(r^2)$, $\Ls(r^2)$, $A^{-1}(p^2)$, and ${\cal M}(p^2)$ are shown in \fig{fig:inputs}.

\section{Changing renormalization schemes}\label{app:renorm}

In this Appendix we show how the lattice inputs given in different renormalization schemes may be converted to the \MOMt, defined in \1eq{MOMtilde}.

We begin by introducing the renormalization constants
\begin{align}
\Delta_{\srm B} &= Z_A \Delta_{\srm R} \,, & F_{\!\srm B} &= Z_c F_{\!\srm R} \,, & A_{\srm B} &= Z_{\rm F}^{-1} A_{\srm R} \,, \nonumber\\
\fatg^\mu_{\!\!c,\,\srm{B}} &= Z_1^{-1}\fatg^\mu_{\!\!c,\,{\srm R}} \,, &\fatg_{\!\!{\srm B}}^\mu &= Z_2^{-1}\fatg^\mu_{\!\!{\srm R}} \,, &\fatg_{\!\!{\srm B}}^{\alpha\mu\nu} &= Z_3^{-1} \fatg_{\!\!{\srm R}}^{\alpha\mu\nu} \,, \nonumber\\
H_{\srm B} &= Z_H^{-1} H_{\srm R} \,, &g_{\srm B} &= Z_g g_{\srm R} \,, &Z_g &= Z_3 Z_A^{-3/2} \,, \label{Zs_def}
\end{align}
where we denote with subscripts ``${\rm B}$'' and ``${\rm R}$'' the bare and renormalized quantities, respectively. Then, the various STIs of the theory lead to the relations~\cite{Marciano:1977su}
\be 
\frac{Z_3}{Z_A} = \frac{Z_1}{Z_c} = \frac{Z_2}{Z_{\rm F}} \,. \label{Z_STIs}
\ee
Moreover, in the Landau gauge, 
we have that $Z_H = Z_1$, 
where $Z_1$ is {\it finite}~\cite{Taylor:1971ff}. From now on, the index ``R'' will be substituted by the notation introduced below, identifying the different renormalization schemes under consideration.

Specifically, the required conversions 
involve three different renormalization schemes, namely:
\begin{enumerate}
\item The \MOMt{} scheme, defined by \1eq{MOMtilde}, which is used in the lattice determination of $\lambda_1(p^2)$. Quantities without any special renormalization index, \eg $\lambda_1(p^2)$, will be understood to be renormalized in the \MOMt{} scheme.

\item The ``asymmetric MOM'' scheme, 
\be 
\Delta_{\srm{asym}}^{-1}(\mu^2) = \mu^2\,, \quad F_{\!\!\srm{asym}}(\mu^2) = 1 \,, \quad A_{\srm{asym}}(\mu^2) = 1 \,, \quad \Ls^{\srm{asym}}(\mu^2) = 1 \,. \label{asym}
\ee
which is specified by the index ``asym''. This scheme was used in~\cite{Aguilar:2019uob} for the lattice determination of the ($N_f = 3$) classical form factor $\Ls(p^2)$ of the three-gluon vertex.

\item The Taylor scheme, which capitalizes on the Landau gauge finiteness of the ghost-gluon vertex~\cite{Taylor:1971ff}, and is defined by the prescription
\be 
\Delta_{\srm T}^{-1}(\mu^2) = \mu^2\,, \quad F_{\srm T}(\mu^2) = 1 \,, \quad A_{\srm T}(\mu^2) = 1 \,, \quad Z^{\srm T}_1 = Z^{\srm T}_H = 1 \,, \label{Taylor}
\ee
where we assign an index ``${\rm T}$'' to quantities renormalized in this scheme. The value of the coupling constant determined in the Taylor scheme in~\cite{Blossier:2010ky} will be used to determine the coupling constants in the above two schemes.

\end{enumerate}

Now, let us note that combining \1eq{Zs_def} with each of the renormalization prescriptions of \3eqs{MOMtilde}{asym}{Taylor} implies
\be 
Z_A = Z_A^{\srm{asym}} = Z_A^{\srm T} = \mu^2 \Delta_{\srm B}(\mu^2) \,, \quad \Longrightarrow\quad \Delta(p^2) = \Delta_{\srm{asym}}(p^2) = \Delta_{\srm T}(p^2) \,,
\ee
since the bare gluon propagator does not depend on the renormalization scheme. Similarly, $F(p^2) = F_{\!\!\srm{asym}}(p^2) = F_{\srm T}(p^2)$ and $A(p^2) = A_{\srm{asym}}(p^2) = A_{\srm T}(p^2)$. Hence, we can suppress the indices specifying renormalization schemes in the propagators. 

On the other hand, the renormalization constant of a given vertex has a different value in each of the above schemes, \eg $Z_3^{\srm{asym}} \neq Z_3$. As it turns out, the ratios between these values, \eg $Z_3^{\srm{asym}}/Z_3$, can be conveniently expressed in terms of three special renormalization group invariant (RGI) effective couplings, defined in terms of the quark-gluon, three-gluon, and ghost-gluon vertices as
\begin{align} 
\alpha_{\s{qg}}(p^2) =&\, \alpha_s^{\srm B} p^2 \Delta_{\srm B}(p^2) [\lambda_1^{\srm B}(p^2)]^2/A^2_{\srm B}(p^2) \,, \nonumber\\
\alpha_{\s{3g}}(p^2) =&\, \alpha_s^{\srm B}[ p^2 \Delta_{\srm B}(p^2)]^3[\Ls^{\srm B}(p^2) ]^2 \,, \nonumber\\
\alpha_{\s{cg}}(p^2) =&\, \alpha_s^{\srm B} p^2 \Delta_{\srm B}(p^2) F^2_{\srm B}(p^2) \,. \label{eff_coup}
\end{align}
Note that the $\alpha_{\s{cg}}(p^2)$ is the well-known Taylor coupling~\cite{Boucaud:2008gn,Aguilar:2009nf,Blossier:2010ky}, which takes advantage of the fact that, in Landau gauge, the unrenormalized ghost-gluon vertex in the soft-ghost configuration reduces to tree-level~\cite{Taylor:1971ff}, \ie $B_1^{\srm B}(r^2,0,r^2) - B_2^{\srm B}(r^2,0,r^2)  = 1$, in the language of \1eq{ghost_gluon_tens}.

\subsection{Taylor to \texorpdfstring{\MOMt}{MOM tilde} conversion and the value of \texorpdfstring{$Z_H$}{ZH}}\label{subsec:Tay_MOMt}

To convert the value of the coupling between the Taylor and \MOMt{} schemes, let us first note that \2eqs{Zs_def}{Z_STIs} imply
\be 
Z_H = Z_1 = Z_c Z_2/Z_{\rm F} \,.
\ee
Then, using \1eq{MOMtilde} to write the renormalization constants in terms of the bare Green's functions, one shows that
\be 
Z_H = \frac{F_{\srm B}(\mu^2)A_{\srm B}(\mu^2)}{\lambda_1^{\srm B}(\mu^2)} = \sqrt{
\frac{\alpha_{\s{cg}}(\mu^2)}{\alpha_{\s{qg}}(\mu^2) }} \,. \label{Momtilde_conversions}
\ee

Now, we can use \1eq{Momtilde_conversions} to determine $\alpha_{\s{qg}}(p^2)$ from the known $\alpha_{\s{cg}}(p^2)$ of~\cite{Blossier:2010ky}. First, at a large momentum scale, here chosen as \mbox{$\nu = 7$~GeV}, we can evaluate \1eq{Momtilde_conversions} at one loop with massless quarks. Using the well-known fact that \mbox{$A_{\srm B}(p^2) = 1$} to one loop in the Landau gauge~\cite{Pascual:1984zb,Muta:1987mz}, together with the corresponding one-loop results for $F_{\!\srm B}(\nu^2)$ and $\lambda_1^{\srm B}(\nu^2)$ given in~\cite{Davydychev:2000rt}, we find 
\be 
Z_H(\nu = 7~\text{GeV}) = 1 + \frac{3\alpha_s(\nu = 7~\text{GeV})}{16\pi} \,.
\ee
Then, using the Taylor scheme value of~\cite{Blossier:2010ky} for \mbox{$\alpha_s(\nu = 7~\text{GeV}) = 0.223$}, we obtain that \mbox{$Z_H(\nu = 7~\text{GeV}) = 1.013$}, and therefore \mbox{$\alpha_{\s{qg}}(\nu = 7~\text{GeV}) = 0.217$}.

Then, since $\alpha_{\s{qg}}(p^2)$ is RGI, we can rewrite \1eq{eff_coup} in terms of quantities renormalized in the \MOMt{} scheme with \mbox{$\mu = 2$~GeV}, \ie
\be 
\alpha_{\s{qg}}(p^2) = \alpha_s p^2 \Delta(p^2) \lambda_1^2(p^2)/A^2(p^2) \,, \label{alpha_qg_MOMt}
\ee
and we determine the corresponding value of 
\be 
\alpha_s(\mu = 2~\text{GeV}) = 0.470(7) \quad [{\rm L08}], \qquad \alpha_s(\mu = 2~\text{GeV}) = 0.471(8) \quad [{\rm L07}], \label{alpha_MOMt}
\ee
by requiring that \1eq{alpha_qg_MOMt} reproduces the previously determined value of $\alpha_{\s{qg}}(\nu = 7~\text{GeV})$.

Lastly, with the full momentum dependence of $\alpha_{\s{qg}}(p^2)$ in hand, we return to \1eq{Momtilde_conversions} to determine 
\be 
Z_H = 1.120(8)\, \quad [{\rm L08}], \qquad Z_H = 1.121(9)\,\quad [{\rm L07}]. 
\label{ZH_val}
\ee

\subsection{Asymmetric MOM to \texorpdfstring{\MOMt}{MOM tilde} conversion}\label{subsec:asym_MOMt}

Now we consider the conversion of $\Ls(p^2)$ from the asymmetric to the \MOMt{} scheme.

To this end, note first that \1eq{Zs_def} implies
\be 
\Ls(p^2) = \frac{Z_3}{Z_3^{\srm{asym}}}  \Ls^{\srm{asym}}(p^2) \,. \label{asym_conversions_step1}
\ee
Then, our task amounts to computing the ratio $Z_3/Z_3^{\srm{asym}}$.

Next, it follows from the renormalization condition of \1eq{asym} that $Z_3^{\srm{asym}} = 1/\Ls^{\srm B}(\mu^2)$, whereas \2eqs{MOMtilde}{Z_STIs} imply
\be 
Z_3 = \frac{\mu^2 \Delta_{\srm B}(\mu^2)A_{\srm B}(\mu^2)}{\lambda_1^{\srm B}(\mu^2)} \,.
\ee
Hence, by combining the above results we obtain
\begin{align} 
\frac{Z_3}{Z_3^{\srm{asym}}} =&\,  \frac{\mu^2 \Delta_{\srm B}(\mu^2)A_{\srm B}(\mu^2)\Ls^{\srm B}(\mu^2)}{\lambda_1^{\srm B}(\mu^2)} = \sqrt{
\frac{\alpha_{\s{3g}}(\mu^2)}{\alpha_{\s{qg}}(\mu^2)} } \,.  
\label{asym_conversions}
\end{align}

Now, to fix the value of $\alpha_{\s{3g}}(\mu^2)$ we use the same strategy used in the determination of \mbox{$\alpha_{\s{qg}}(p^2)$}. 

At a large momentum scale, \mbox{$\nu = 7$~GeV}, we can evaluate the combination of bare Green's functions at one loop. Using the textbook result for $\Delta_{\srm B}(\nu^2)$ and \mbox{$A_{\srm B}(\nu^2) = 1$}~\cite{Pascual:1984zb,Muta:1987mz} and the results of \cite{Davydychev:1996pb,Davydychev:2000rt} for the vertices we obtain
\be 
\left. \frac{Z_3}{Z_3^{\srm{asym}}}\right\vert_{\nu = 7~\text{GeV}}\!\!\! = 1 + \frac{\alpha_s(\nu = 7~\text{GeV})}{ 96\pi }( 129 - 16 N_f ) = 1.072 \,,
\ee
which implies \mbox{$\alpha_{\s{3g}}(\nu = 7~\text{GeV}) = 0.249$}.

Then, from the RGI nature of $\alpha_{\s{3g}}(p^2)$, we can compute its entire momentum dependence using ingredients renormalized in the asymmetric scheme. In particular, we obtain the three-gluon coupling \mbox{$\alpha_{\s{3g}}(\mu = 2~\text{GeV}) = 0.635$}, and
\be 
\frac{Z_3}{Z_3^{\srm{asym}}} = 1.16 \,. \label{Z3conv}
\ee
%

\section{Unquenched ghost-gluon vertex}\label{app:ghost-gluon}

In this Appendix we briefly describe the determination of the classical form factor $B_1(r^2,k^2,q^2)$ of the ghost-gluon vertex [see \1eq{ghost_gluon_tens}], which is relevant for the calculation of the quark-ghost form factors, $K_i$.

The one-loop dressed SDE of the ghost-gluon vertex is formally identical to its quenched version, given diagrammatically in Fig.~12 of~\cite{Ferreira:2023fva}; 
the dependence on $N_f$
enters through the various 
quantities 
comprising the diagrams, such as  
the coupling, the gluon and ghost propagators, and the three-gluon vertex.

For the three-gluon vertex, we use the approximation given in \1eq{planar}. Then, an appropriate tensor projection of the SDE yields a dynamical equation for $B_1(r^2,k^2,q^2)$, which is given by Eq.~(88) of~\cite{Ferreira:2023fva}. Finally, this equation is solved numerically, using \1eq{alpha_MOMt} for $\alpha_s$, the L08 setup value given in \1eq{ZH_val} for $Z_1 = Z_H$, and the fits given by \1eq{L_fit} for $\Delta(r^2)$, $F(r^2)$ and $\Ls(r^2)$.

To expedite the comparison between \mbox{$N_f = 2$} and quenched results for the ghost-gluon vertex, we will show its classical form factor renormalized in the Taylor scheme, $B_1^{\srm T}$. Its \MOMt{} value, $B_1$, is obtained immediately through
\be 
B_1(r^2,k^2,q^2) = Z_H B_1^{\srm T}(r^2,k^2,q^2) \,,
\ee
where $Z_H$ is given by \1eq{ZH_val}.

\begin{figure}[ht]
\includegraphics[width=0.45\linewidth]{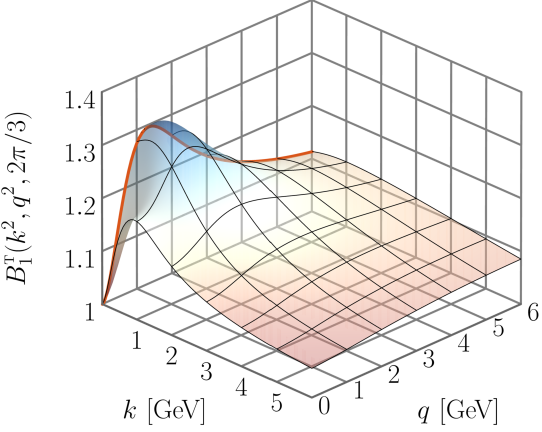} \hfil 
\includegraphics[width=0.45\linewidth]{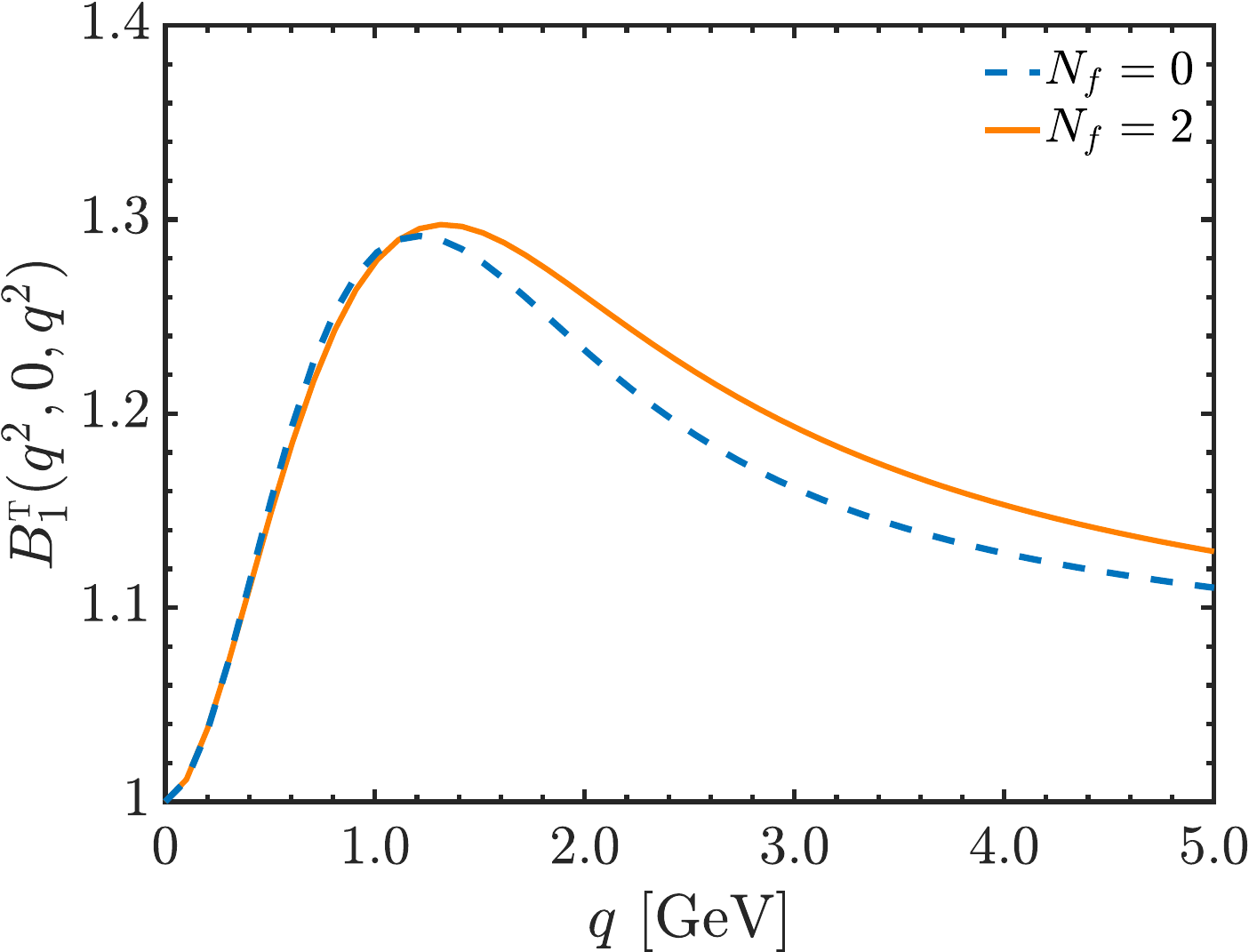}
\caption{  Left: Classical form factor, $B_1^{\srm T}(r^2,k^2,q^2)$, of the ghost-gluon vertex renormalized in the Taylor scheme, for a fixed value \mbox{$\phi = 2\pi/3$} of the angle between $k$ and $q$. The orange continuous line highlights the soft-ghost ($k = 0$) limit, respectively. Right: Comparison between the \mbox{$N_f = 2$} (orange continuous) and quenched (blue dashed) values of the soft-ghost limit, $B_1^{\srm T}(q^2,0,q^2)$.}
\label{fig:B1_SDE}
\end{figure}

On the left panel of \fig{fig:B1_SDE} we show $B_1^{\srm T}(r^2,k^2,q^2)$ as a function of the magnitudes of the ghost and gluon momenta,
$k$ and $q$, respectively; 
for the angle $\phi$ between them we choose the representative value \mbox{$\phi = 2\pi/3$}. On this panel we note that $B_1^{\srm T}(r^2,k^2,q^2)$ has the same qualitative behavior as its quenched form obtained in various studies\footnote{For related studies within the Curci-Ferrari model, see~\cite{Mintz:2017qri,Barrios:2020ubx}.}~\mbox{\cite{Schleifenbaum:2004id,Ilgenfritz:2006he,Aguilar:2013xqa,Cyrol:2016tym,Huber:2018ned,Aguilar:2018csq,Huber:2020keu,Aguilar:2021okw}}. In particular,  it is nearly identical in shape to the result shown in Fig. 13 of~\cite{Ferreira:2023fva}, which uses the same approximation of \1eq{planar} for the three-gluon vertex. 

Next, we focus on the soft-ghost limit ($k = 0$), which appears as an ingredient in \1eq{Ki_SDE} for the $K_i(p^2)$. This limit corresponds to the slice highlighted as an orange continuous curve on the left panel of \fig{fig:B1_SDE}. On the right panel of the same figure, we compare the $N_f = 2$ value of $B_1^{\srm{T}}(q^2,0,q^2)$ to its quenched counterpart computed in~\cite{Ferreira:2023fva}, represented here as a blue dashed line. The results are found to differ by less than $2.75\%$ within the entire momentum range, and are virtually indistinguishable for $q < 1$~GeV. Their most noticeable difference is in the ultraviolet, where the $N_f = 2$ result is seen to be systematically larger than its quenched version. The same pattern is found for all kinematic configurations.

Note that the observed enhancement of the unquenched $B_1^{\srm{T}}(q^2,0,q^2)$ in the ultraviolet is compatible with 
perturbation theory. Indeed, a one-loop calculation yields
\be 
B_1^{\srm{T}}(q^2,0,q^2) = 1 + \frac{33\alpha_s^{\srm{T}}}{32\pi} \,,
\ee
which depends on $N_f$ only through the value of $\alpha_s^{\srm{T}}$. Then, since $\alpha_s^{\srm{T}}$ increases with $N_f$~\cite{Ayala:2012pb}, so does $B_1^{\srm{T}}(q^2,0,q^2)$.


%

\end{document}